\documentclass[10pt]{article}
\usepackage{amssymb}
\usepackage{amsmath}
\usepackage[all]{xy}
\usepackage{amsmath}
\usepackage{amsfonts}
\usepackage{amssymb}
\usepackage{amscd}
\usepackage{amsthm}
\usepackage{latexsym}
\usepackage{amsbsy}
\usepackage{graphicx}
\usepackage{float}
\usepackage[toc,page]{appendix}
\usepackage{caption}
\usepackage{subcaption}
\usepackage{epstopdf}
\usepackage{authblk}
\usepackage{amssymb,latexsym,amsmath,amscd,slashed,mathtools,mathrsfs,bm,stmaryrd,dsfont,amsthm}
\usepackage{setspace,scalerel,color,enumerate}
\usepackage[multiple]{footmisc}
\usepackage{authblk}
\usepackage[toc,page]{appendix}

\newtheorem{definition}{Definition}[section]

\DeclareMathOperator{\tr}{Tr}
\DeclareMathOperator{\End}{End}

\newcommand{\ci}{\mathrm{i}}

\usepackage{tocloft}

\usepackage{hyperref}
\hypersetup{
	colorlinks,
	citecolor=black,
	filecolor=black,
	linkcolor=black,
	urlcolor=black
}
\makeatletter
\def\section{\@startsection{section}{1}{\z@}{-3.25ex plus -1ex minus
		-.2ex}{1.5ex plus .2ex}{\normalfont\bfseries}}
\def\subsection{\@startsection{subsection}{1}{\z@}{-3.25ex plus -1ex
		minus -.2ex}{1.5ex plus .2ex}{\normalfont\itshape}}
\date{}
\makeatother
\title{From Noncommutative Geometry to Random Matrix Theory}
\author{Hamed Hessam, Masoud Khalkhali, Nathan Pagliaroli, and Luuk S. Verhoeven}

\affil{Department of Mathematics, Western University\\
	London, Ontario, Canada\footnote{\emph{Email addresses}: hhessam@uwo.ca, masoud@uwo.ca, npagliar@uwo.ca, lverhoe@uwo.ca}}

\usepackage{graphicx}
	\graphicspath{{./Diagrams/Type(1,0)/}{./Diagrams/Type(2,0)/}}

\begin{document}

	\maketitle
	\begin{abstract}
 We review recent progress in the analytic study of random matrix models suggested by noncommutative geometry. One considers fuzzy spectral triples where the space of possible Dirac operators is assigned a probability distribution. These ensembles of Dirac operators are constructed as toy models of Euclidean quantum gravity on finite noncommutative spaces and display many interesting properties. The ensembles exhibit spectral phase transitions, and near these phase transitions they show manifold-like behavior. In certain cases one can recover Liouville quantum gravity in the double scaling limit.  We highlight examples where bootstrap techniques, Coulomb gas methods, and Topological Recursion are applicable. 
	\end{abstract}


\tableofcontents

\section*{Introduction}

In this paper we would like to give an overview of some of the recent developments on the intersection of noncommutative geometry, random matrix theory, and Euclidean quantum gravity. The existence of the Planck length puts restrictions on the nature of a theory of spacetime suitable for a quantum theory of gravity. In fact,  a combination of the Heisenberg uncertainty principle  and Einstein's general relativity shows that, due to the formation of black holes in small length scales, spacetime cannot be a smooth manifold. There have been several possibilities suggested  for a replacement of classical spacetime. A noncommutative space, in the sense of spectral triples, is one such proposal \cite{Connes-Hilbert}. Other possibilities include spin networks \cite{spin networks}, random tensors \cite{random tensors}, spin foams \cite{spin foam}, and loop quantum gravity \cite{loop quantum gravity}.  If the spectral triple is assumed to be finite one can use computer simulations and random matrix theory techniques to explore these models in detail.

Random matrices first appeared in physics in a series of papers by Wigner in the mid 50's \cite{Wigner 1,Wigner 2,Wigner 3}, where Wigner modeled the Hamiltonians of heavy nuclei by large random Hermitian or symmetric matrices. Since then random matrices have found applications in many other areas of physics, in particular in models of two dimensional quantum gravity.
This was first seen in the matrix integral used by Kontsevich to prove Witten's conjecture  \cite{Witten, Kontsevich}. Around the same time it was found that artifacts of two dimensional conformal field theory coupled with gravity, sometimes called Liouville quantum gravity (LQG), could be obtained from certain matrix models in the double scaling limit \cite{DS multi 1,Ds multi 2,Eynard LQG}. More recently it was discovered in \cite{JT 1,JT 2,JT 3,JT 4} that the genus expansion of partition functions in Jackiw–Teitelboim (JT) gravity can be computed using random matrix techniques. In particular, a process known as Topological Recursion, original developed in \cite{Eynard TR}, was applied. In fact Topological Recursion can be used in all the cases stated above. We will discuss this method in Appendix \ref{SDE's and TR}.

We are specifically interested in toy models of Euclidean quantum gravity where integration over metrics is replaced by integration over Dirac operators on a fixed finite noncommutative space, as proposed by Barrett and Glaser \cite{Barrett2016}. This scenario quickly leads to very interesting  multi-trace multi-matrix models. As a rule such models are hard to analyze using standard methods. Yet the fact that they are obtained from specific potentials defined on the space of Dirac operators gives  them a special structure and hence the possibility of analytic study. 

The partition function of these models is of the form
\begin{equation*}
	Z=\int_{\mathcal{D}}e^{-S(D)}dD,
\end{equation*}
where the integration is carried out over the space $\mathcal{D}$ of Dirac operators. This can be justified by general principles of noncommutative geometry, starting from  the fact that the metric structure of a smooth spin manifold is encoded in its Dirac operator \cite{Connes95,Connes-2013,QFTNCG, Marcolli}.  In particular Dirac operators are  taken as  dynamical variables and play the role of metric  fields in gravity. 

Such a matrix integral is not necessarily convergent, nor does it require a real valued action $S$. However, they may always be interpreted as formal matrix integrals, which are the generating functions of certain types of maps \cite{BPZ,Eynard2018}.
The action functional $S$ is, for models considered in this survey, alwas chosen in such a way that the partition function $Z$ is absolutely convergent and finite. For example, we can choose $ S (D)= \text{Tr} (f (D))$ for a  real polynomial $f$ of even degree with a positive leading coefficient. 
For more details on  formal and convergent matrix models see \cite{Eynard combin, Deift}.

There are several benefits Dirac ensembles have over the usual random matrix ensembles. A random matrix model can be interpreted as a zero-dimensional quantum field theory  or, if the model is formal, it may be viewed as a discretized path integral from string theory, where maps act as discrete surfaces. A Dirac ensemble maintains these interpretations while also being formalized as a computable noncommutative path integral over metrics, represented as Dirac operators, which is a key feature of a theory of quantum gravity. 
Additionally Dirac ensembles have an interpretation as a random noncommutative space. The probability distribution on Dirac operators corresponds to a probability distribution on noncommutative geometries. This idea is first mentioned in \cite{Barrett2016} and explored in detail in \cite{Spectral estimators}, where geometric quantities are determined using the spectrum. 

It is worth noting that the connections between matrix integrals, noncommutative geometry, and physics do not start with Dirac ensembles. In \cite{path integral finite ncg} path integrals over finite spectral triples are also studied. Additionally the Kontsevich model and some of its generalizations appear in noncommutative quantum field theory \cite{Quartic K 1,Quartic K 2, Quart K 3, Quartic K 4}. In particular, it is conjectured that a quartic version of the Kontsevich model obeys a generalized version of Topological Recursion known as Blobbed Topological Recursion \cite{blobbed1,blobbed2}. For a review of these developments see \cite{Quartic K review}.

For Dirac ensembles a connection was recently established with LQG \cite{HKP 2}. The authors hope that one day a connection to JT gravity will also be established using similar methods. We will discuss this development and the relevant background.

This paper is organized as follows. In Section 1 we define Dirac ensembles, give examples of how they can be studied and how they give rise to interesting theories, such as Yang-Mills-Higgs fuzzy spaces.
In Section 2 we review results on the spectral distributions and phase transitions in Dirac ensembles.
In Section 3 we recall the bootstrap method as applied to Dirac ensembles.
In Section 4 we formulate some natural open problems and directions for further study. Finally, in the Appendices we recall some aspects of random matrix theory, the Schwinger-Dyson equations and Topological Recursion.

Acknowledgments: We would like to deeply thank the referees whose careful reading of the text and many  constructive suggestions led to an improved version of this paper.

\section{Random matrix models from spectral triples}

In this section, before we introduce the key concept of a {\it Dirac ensemble}, we shall first  briefly discuss the notions of  {\it spectral triples}  and in particular {\it  fuzzy spectral triples} which are the base for the Dirac ensembles appearing in this survey. It may be helpful for readers unfamiliar with random matrix ensembles to review Appendices \ref{convergent integrals} and \ref{formal matrix models}.

\subsection{Fuzzy spectral triples}

{\it Spectral triples} were introduced by Connes in \cite{NCG and Reality} (see also \cite{QFTNCG}) and are defined by data $(\mathcal{A},\mathcal{H}, D)$ where $\mathcal{A}$ is a unital, involutive, complex, associative algebra acting by bounded operators on a complex Hilbert space $\mathcal{H}$, and  $D$ is a self-adjoint (in general unbounded) operator acting on $\mathcal{H}.$ 
This data is further required to satisfy certain finiteness and regularity conditions, which are automatically satisfied if $\mathcal{A}$ and $\mathcal{H}$ are finite dimensional.
Since this will always be the case in this survey we omit these conditions here.

A {\it real spectral triple}  is a spectral triple  equipped with two additional operators $J$ and $\gamma$ called the {\it charge conjugation} and the {\it chirality operator}, where  $J: \mathcal{H} \to \mathcal{H}$ is an anti-linear real structure, with the requirement that $[a, JbJ^{-1}]=0$ for all $a, b \in \mathcal{A}$, and $\gamma: \mathcal{H} \to \mathcal{H}$ is a self-adjoint operator with $\gamma^2 = 1$. The data $(\mathcal{A},\mathcal{H}, D, J, \gamma)$ is required to satisfy some further compatibility conditions between $D$, $J$, $\gamma$, and the representation of $\mathcal{A}$ which we will recall in Definition \ref{fuzzy def}.

The overall idea is that a real spectral triple is a noncommutative analogue of a $\text{spin}^c$  Riemannian manifold and its canonical  Dirac operator. In fact, any closed $\text{spin}^c$ Riemannian manifold $M$ defines a real (commutative) spectral triple as follows. The algebra $\mathcal{A}= C^{\infty} (M)$ is the algebra of smooth complex valued functions on $M$. The Hilbert space consists of square integrable sections of the spinor bundle, with $\mathcal{A}$ acting as multiplication operators. The operator  $D$ is the Dirac operator of $M$ acting on the spinors, and $J$ and $\gamma$ are the standard charge conjugation and chirality operators. The reconstruction theorem of Connnes states that, conversely, a commutative real spectral triple, i.e. a triple where $\mathcal{A}$ is commutative, satisfying some natural conditions is the spectral triple of a $\text{spin}^c$ Riemannian manifold \cite{Connes-2013}. The reader can find further details and many interesting commutative and non-commutative examples of spectral triples in the book of Connes and Marcolli \cite{QFTNCG}, as well as their applications to the standard model of elementary particles and quantum field theory in general.  

A spectral triple is called {\it finite} if both $\mathcal{A}$ and $\mathcal{H}$ are finite dimensional vector spaces. In this paper we shall primarily consider a subclass of finite real spectral triples called {\it fuzzy spectral triples} or {\it fuzzy geometries} introduced and classified in their present form by Barrett in \cite{Barrett2015}. These should be thought of as $\text{spin}^c$  Riemannian manifolds with a finite resolution or Plank length. 
It should be noted that important examples of fuzzy spectral triples like the fuzzy sphere \cite{fuzzy sphere, fuzzy sphere spec, fuzzy sphere grosse, Barrett2015, fuzzy sphere Torres} and fuzzy tori \cite{Fuzzy Torus Large N matrices,Fuzzy torus generalized,Fuzzy torus as spectral triple,Fuzzy Torus convergence} were defined and studied for their own interest before the concept of fuzzy geometry was coined.
We should also mention that finite dimensional real spectral triples have been fully classified by Krajewski in \cite{Finite spectral triples}. Further references include \cite{QFTNCG, Marcolli, Barrett2015, VS}. 

We will now specialize further to the class of fuzzy spectral triples.
Let ${{\mathrm{C \ell}}_{p,q}}$, for non-negative integers $p$ and  $q$, denote the real Clifford algebra associated to the vector space ${\mathbb{R}^n}$, $n=p+q$, and the pseudo-Euclidean metric ${\eta}$ of signature ${(p,q)}$ given by 
\begin{equation*} 
  {\eta} (v,v) = {{v_1}^2} + \cdots + {{v_p}^2} - {{v_{p+1}}^2} - \cdots -  {{v_{p+q}}^2} \,
 	, \quad v \in \mathbb{R}^n \, .
\end{equation*}
Let ${ {\mathbb{C} \ell}_n \coloneqq {\mathrm{C \ell}}_{p,q} {\otimes_{\mathbb{R}}} \mathbb{C}\,}$ denote the  complexification of  ${{\mathrm{C \ell}}_{p,q}}$. Let ${ {\{ e_i \}}_{i=1}^n }$  denote the  standard  basis of ${\mathbb{R}^n}$. The \emph{chirality element} $\Gamma \in { {\mathbb{C} \ell}_n } $ is defined  by
\begin{equation*}
  \Gamma = \ci^{\frac{1}{2} s (s+1)} \, e_1 e_2 \cdots e_n \, ,
\end{equation*}
where ${s \equiv q-p \pmod{8}}$ is known as the $KO$-dimension. 
We denote by ${V_{p,q}}$ the unique (up to unitary equivalence) hermitian irreducible  ${{\mathrm{C \ell}}_{p,q} \,}$-module, where for $n=p+q$ odd the chirality element $\Gamma$ acts trivially on $V_{p,q}\,$. 
The module $V_{p, q}$ also comes with a charge conjugation operator  ${C : V_{p,q} \to V_{p,q}}$ (see  \cite{Marcolli,Barrett2015} for details).
 
\begin{definition}\label{fuzzy def}
  A  {\emph fuzzy spectral triple of type, or signature, ${(p,q)}$} is a finite real spectral triple
  ${\left( { \mathcal{A}, \mathcal{H} , D , J, \gamma } \right)}$ where
  \begin{itemize}
 		\item
 		${ \mathcal{A}= {{\mathrm{M}}_N (\mathbb{C})} }$  is the algebra of complex $N \times N$ matrices,
 		\item
 		${ \mathcal{H} = V_{p,q} \otimes {{\mathrm{M}}_N (\mathbb{C})} }$ with the inner product
 		$${ \langle {u \otimes A , v \otimes B} \rangle = \langle u , v \rangle \, \tr \left( A B^\ast \right)
 			\, , \quad u,v \in V_{p,q} \, , \  A,B \in {{\mathrm{M}}_N (\mathbb{C})} },$$
 		\item The action of $\mathcal{A}$ on $\mathcal{H}$ is defined by 
 		${ A \cdot (v \otimes B) = v \otimes \left( AB \right),   }$
 			\item The charge conjugation operator is 
 		${ J (v \otimes A) = (C v) \otimes A^\ast  \,}$,
 		\item The chirality operator is defined as 
 		${ \gamma (v \otimes A) = (\Gamma v) \otimes A },$
 	
 		\item The Dirac operator $D$ satisfies:\\
 		a) $D^* =D,$\\
 		b) $D\gamma = - (-1)^s \gamma D,$\\
 		c) $DJ= \epsilon' JD$ , where $\epsilon'  = 1$ for $s = 0, 2, 3, 4, 6, 7$  and $\epsilon' = -1$ for $s = 1$ or $ 5$,\\
 		d) $[[D, a], JbJ^{-1}]=  0$ for all $a, b \in \mathcal{A}.$ 
 	\end{itemize}
  The quantity $n = p+q$ is called the dimension of the fuzzy spectral triple, the quantity $s = q- p$ is the $KO$-dimension.
\end{definition}
 
The main benefit of considering fuzzy spectral triples is that their Dirac operators can be expressed in terms of the gamma matrices ${\gamma^i \,}$ (the image of $e_{i}$ in the Clifford algebra), and commutators or anti-commutators with  Hermitian or skew-Hermitian matrices. More precisely, Barrett proved in \cite{Barrett2015} that the Dirac operator of a fuzzy spectral triple is always of the form
\begin{equation}
	D=\sum \gamma^{I}\otimes \{K_{I},\cdot\}_{e_{I}}
  \label{general fuzzy Dirac}
\end{equation}
where the sum is over increasingly ordered multi-indices $I$.
If $\gamma^I$ is Hermitian, $e_I = 1$ and $\{K_{I},\cdot\}_{e_{I}} = \{H_{I},\cdot \}$, where $H_{I}$ is some Hermitian matrix.
If $\gamma^I$ is skew-Hermitian, $e_I = -1$ and $\{K_{I},\cdot\}_{e_{I}} = [L_{I},\cdot ]$, where $L_{I}\,$is some skew-Hermitian matrix. 
This allows us to effectively parametrize the space of Dirac operators by matrices.

One prominent example of a fuzzy spectral triple is the fuzzy sphere \cite{Stanovich, The moyal representation for spin,fuzzy sphere, fuzzy sphere grosse, fuzzy sphere spec}.
Let $J_1, J_2, J_3$ be the standard skew-Hermitian generators of $\mathfrak{su}_2$ and denote the $2j+1$-dimensional irreducible representation of $\mathfrak{su}_2$ by $(\pi_j, V_j)$.
An initial definition of a Dirac operator for the fuzzy sphere is then defined on $\mathbb{C}^2 \otimes M_{2j+1}(\mathbb{C}) \cong \mathbb{C}^2 \otimes V_j \otimes V_j^*$ by
$$
  d = 1 \otimes 1 + \gamma^\mu \otimes \left[\pi_j(J_\mu), \cdot\right]
$$
with $\gamma^{\mu} = i\sigma^\mu$ for the Pauli matrices $\sigma^{\mu}$.
This definition has two problems, it is of signature $(0, 3)$ which has $KO$-dimension 3 rather than the desired 2, and it does not admit a grading \cite{fuzzy sphere spec}.
Hence $d$ cannot be the Dirac operator for a fuzzy geometry.
Instead we form a Dirac operator of signature $(1, 3)$, by $D = \sigma^1 \otimes d$ which admits a grading $\sigma^3 \otimes 1$.
The real structure is given by $J(v \otimes a) = \sigma_2 \overline{v} \otimes a^*$.

\subsection{Dirac ensembles}

By a {\it Dirac ensemble}  we mean a statistical ensemble of {\it  fuzzy  spectral triples} $(\mathcal{A},\mathcal{H}, D, J, \gamma)$  where the ``Fermion space''  $(\mathcal{A},\mathcal{H}, J, \gamma)$, is kept fixed but the Dirac operator $D$ is a random variable with a given probability density. 
This a non-commutative analogue to a probability distribution on the space of metrics on a given manifold, as the Dirac operator encodes the metric structure while the algebra encodes the topology.

More precisely, let $\mathcal{D}$ denote the set of all possible  Dirac operators $D:\mathcal{H} \rightarrow \mathcal{H}$ such that the quintuple $(\mathcal{A},\mathcal{H}, D, J, \gamma)$ satisfies the conditions a), b), c), and d) of definition \ref{fuzzy def}. Clearly
$\mathcal{D} \subset \text{End} (\mathcal{H})$ is a  real  subspace,  hence it is equipped with an inner product and thus a natural Lebesgue measure which we denote by $dD$. 
Given a choice of action functional $S:\mathcal{D} \to \mathbb{R}$, usually a polynomial whose choice is part of the data for a Dirac ensemble, the probability density on $\mathcal{D}$ is defined by
$$
	\frac{1}{Z}e^{-\tr S(D)} dD
$$
where
$$
	Z = \int_{\mathcal{D}} e^{-\tr S(D)}dD
$$
is the partition function of the model.


For fuzzy spectral triples the space of Dirac operators can be paramatrized by a combination of (skew)-Hermitian matrices or, by writing the skew-Hermitian matrices as $\ci$ multiplied by a Hermitian one, purely by Hermitian matrices using equation (\ref{general fuzzy Dirac}). 
The probability density, and thus partition function, may then be written as a matrix integral of the form  
\begin{equation} \label{Dirac to matrix}
  Z = \int_{\mathcal{D}} e^{-\tr S(D)}dD = \int_{\mathcal{H}_{N}^{m}} e^{-\tilde{S}( H_{1},  H_{2},..., H_{m})} dH_{1}... dH_{m}, 
\end{equation} 
where $\mathcal{H}_{N}$ is the space of Hermitian $N$ by $N$  matrices. For a given Dirac ensemble defined by $e^{-\tr S(D)}dD$, we refer to the third term in equation (\ref{Dirac to matrix}) as the associated random matrix ensemble.  We will see this correspondence in detail in Sections \ref{One dimensional Dirac} and \ref{two dim Dirac}.


The matrix ensembles associated to Dirac ensembles are usually both multi-trace or multi-matrix.
Since most of the results in random matrix theory are for single matrix and single trace models, this hints that the analytic study of Dirac ensembles as matrix integrals is quite difficult in general. 


 \subsection{One dimensional Dirac ensembles}\label{One dimensional Dirac}
 For fuzzy spectral triples of dimension $n=1$ the only possible signatures are $(1, 0)$ and $(0, 1)$.  In both  case  $ \mathcal{H} = M_N(\mathbb{C})$ and the  two possible choices for $D$ are as follows \cite{Barrett2015, Barrett2016}:
\begin{description}
  \item[Type (1, 0)] The Dirac operator is the anticommutator with a Hermitian matrix  $H$,
    \begin{equation*}
      D = \{H,\cdot\}.
    \end{equation*} The trace of powers of $D$ can be computed by
    \begin{equation*}
      \text{Tr} \, D^{\ell}  = \sum_{k=0}^{\ell} {\ell \choose k}\tr H^{\ell-k}\tr H^{k}.
    \end{equation*}
  \item[Type (0, 1)] The Dirac operator is the commutator with a skew-Hermitian matrix $L$,
    \begin{equation*}
      D =  i [L,\cdot].
    \end{equation*}
      The trace of powers of $D$ can be computed by
    \begin{equation*}
      \text{Tr} \, D^{\ell}  =  \sum_{k=0}^{\ell} {\ell\choose k}(-1)^{k}\tr L^{\ell-k}\tr L^{k}.
    \end{equation*}
\end{description}

For the action functional we can consider a quartic potential 
\begin{equation*}
  Z  = \int_{\mathcal{D}} e^{- g \tr D^{2} - \tr D^{4}} dD,
\end{equation*}
where the real parameter $g$ is called a coupling constant. In type $(1,0)$ the integral is over the space $\mathcal{H}_{N}$ of Hermitian $N \times N$ matrices and the partition function in terms of $H$ is 
\begin{equation*}
	Z = \int_{\mathcal{D}}e^{-g \tr D^2 - \tr D^4 }dD = \int_{\mathcal{H}_{N}}e^{-2N(g \tr H^2 +   \tr H^4) -2g (\tr H)^2 - 8 \,\tr H \tr H^3 - 6(\tr H^2)^2 }dH.
\end{equation*}
The substitution of $\mathcal{H}_N$ for $\mathcal{D}$ is justified as the parametrization of Dirac operators by the Hermitian matrix $H$ is bijective.

In type $(0,1)$ the integral is over the space $\mathcal{L}_{N}$ of skew-Hermitian $N\times N$ matrices, and the partition function is given by 
\begin{align*}
	Z = \int_{\mathcal{D}}e^{-g \tr D^2 - \tr D^4 }dD = \int_{\mathcal{L}_{N}}e^{-2g(N \tr L^{2} - (\tr L)^{2}) - \left( 2N \tr L^{4} -8 \tr L \tr L^{3} + 6 (\tr L^{2})^{2}\right)} dL.
\end{align*}
Note that the kernel of the map $\mathcal{L}_N \to \mathcal{D}$ consists of the scalar matrices, which has lebesgue measure zero, justifying the substituion.
We can write  $L =  iH $, for a  Hermitian matrix $H$, to get 
\begin{align*}
	Z = \int_{\mathcal{D}}e^{-g \tr D^2 - \tr D^4 }dD = i\int_{\mathcal{H}_{N}}e^{2 g (N \tr H^{2} - (\tr H)^{2}) +  \left(- 2N \tr H^{4} + 8 \tr H \tr H^{3} - 6(\tr H^{2})^{2}\right)}dH.
\end{align*}

In \cite{First paper} it was shown that the two terms $(\tr H)^2$ and $(\tr H)\tr H^3$ contribute nothing in the large $N$ limit, giving us the same matrix integral as the above quartic type $(1,0)$ up to a factor of $i$. This idea extends to all type $(1,0)$ and $(0,1)$ Dirac ensembles with even potentials, that is, all  such ensembles will have identical real eigenvalue density in the large $N$ limit. 

If one considers formal Dirac ensembles of types $(1,0)$ or $(0,1)$, it can be shown \cite{AK} that their partition functions and moments are the  generating functions that count combinatorial objects know as stuffed maps in the sense of the work Borot and Shadrin in \cite{blobbed1,blobbed2}. Similar to the more common types of maps that arise in Hermitian matrix ensembles, the matrix integrals that generate stuffed maps obey a generalized form of Topological Recursion, called Blobbed Topological Recursion. Given the genus zero one-point and two-point generating functions one can recursively compute all higher order corrections. In particular, this (blobbed) topological recursion applies to $(1,0)$ or $(0,1)$  Dirac  ensembles with multi-tracial potentials and was studied in \cite{AK}.

\subsection{Two dimensional Dirac ensembles}\label{two dim Dirac}

For two dimensional fuzzy geometries there are three options. In this case the Hilbert space $ \mathcal{H} = \mathbb{C}^{2} \otimes M_N(\mathbb{C})$, where for $p+q=2$, $\mathbb{C}^{2} \cong V_{p,q}$  is the space of spinors. The structure of the Dirac operator depends on the type as follows: 
\begin{description}
  \item[Type (2, 0)] Let \begin{equation*}
  \gamma^{1} = \begin{pmatrix}
  1 & 0 \\
  0 & -1
  \end{pmatrix}, 
  \quad \quad \gamma^{2} = \begin{pmatrix}
  0 & 1 \\
  1 & 0
  \end{pmatrix}.
  \end{equation*}
  Then, 
  \begin{equation*}
  D = \gamma^{1}\otimes \{H_{1},\cdot\} + \gamma^{2} \otimes \{H_{2},\cdot\},
  \end{equation*}
  where $H_1$ and $H_2$ are Hermitian matrices. 
  \item[Type (1, 1)] Let \begin{equation*}
  \gamma^{1} = \begin{pmatrix}
  1 & 0 \\
  0 & -1
  \end{pmatrix},
  \quad \quad \gamma^{2} = \begin{pmatrix}
  0 & 1 \\
  -1 & 0
  \end{pmatrix}.
  \end{equation*}
  Then,  
  \begin{equation*}
  D = \gamma^{1}\otimes \{H,\cdot\} + \gamma^{2} \otimes [L,\cdot],
  \end{equation*}
  where $H$ is Hermitian and $L$ is skew-Hermitian.
  \item[Type (0, 2)] Let \begin{equation*}
  \gamma^{1} = \begin{pmatrix}
  i & 0 \\
  0 & -i
  \end{pmatrix},
  \quad \quad \gamma^{2} = \begin{pmatrix}
  0 & 1 \\
  -1 & 0
  \end{pmatrix}.
  \end{equation*}
  Then, 
  \begin{equation*}
  D = \gamma^{1}\otimes [L_{1},\cdot] + \gamma^{2} \otimes [L_{2},\cdot],
  \end{equation*}
  where $L_1, L_2$ are both skew-Hermitian.
\end{description}

The general structure of trace powers of Dirac operators ensembles of signature above 1 becomes quite complicated and has no obvious patterns. 
They were first studied in \cite{Sanchez}. 

The associated matrix models also rise in complexity, take for example a type $(2,0)$ quartic potential 

\begin{align*}
	Z &= \int_{\mathcal{D}}e^{-\frac{t_{2}}{8}\tr D^2 -\frac{t_{4}}{16}\tr D^{4}}dD\\
	&=\int_{\mathcal{H}_{N}^{2}}e^{\tilde{S}(H_{1},H_{2})}dH_{1}dH_{2},
\end{align*}
where 
\begin{align}
\begin{split}\label{quartic (2,0)}
	\tilde{S}(H_{1},H_{2}) &= -t_{2}(N\tr H_1^2 + N\tr H_2^2) -t_{4}\left(\frac{1}{4}N \tr H_1^4 + \frac{1}{4}N \tr H_2^4  + N\tr H_1^2H_2^2 \right. \\
	 & \left. -\frac{1}{2}N \tr H_1H_2H_1H_2
	+ \frac{3}{4}  (\tr H_1^2)^2  + \frac{3}{4} (\tr H_2^2)^2 + \frac{1}{2} \tr H_1^2 \tr H_2^2\right)
\end{split}
\end{align}
are the contributing terms in the large $N$ limit. This ensemble is a bi-tracial two-matrix model. There are no known applicable analytic techniques from random matrix theory.

Note that similarly to the type $(0,1)$ geometry, Dirac ensembles with skew-Hermitian matrices can always be converted to ensembles of Hermitian matrices. So, in particular, studying models from the above geometries  amounts to  solving  Hermitian two-matrix models.

\subsection{Yang-Mills-Higgs Dirac ensembles}

Dirac ensembles as defined so far describe only the metric structure of a fuzzy space. The space $V_{p, q}$ plays the role of a spinor space and $M_N(\mathbb{C})$ plays the role of $L^2$-functions on the manifold so that together they make a (trivial) spinor bundle.
In order to include a gauge sector we can consider Yang-Mills-Higgs fuzzy spaces \cite{Sanchez3}.
This approach is based on gauge theory on almost commutative manifolds \cite{Spectral action, Neutrino Mixing} (see also chapter 8 of \cite{VS}), and consists of introducing a finite spectral triple playing the role of an additional (trivial) vector bundle which will carry an analogue of a connection.

Concretely, let $M_f = (M_N(\mathbb{C}), \mathcal{H}_N, D_f, J, \gamma)$ be a fuzzy spectral triple in the sense of Definition \ref{fuzzy def} and let $F = (\mathcal{A}_F, \mathcal{H}_F, D_F, J_F, \gamma_F)$ be a finite spectral triple.
This second spectral triple will be referred to as the \textit{gauge} or \textit{finite} spectral triple.
If $\mathcal{A}_F = M_n(\mathbb{C})$ and $\mathcal{H}_F = M_n(\mathbb{C})$ the gauge triple is referred to as a Yang-Mills triple.
The Gauge-Higgs fuzzy space, or Yang-Mills-Higgs fuzzy space if the gauge triple is of Yang-Mills type, is then the product spectral triple $M_f \times F$  given by
$$
  \left(M_N(\mathbb{C}) \otimes \mathcal{A}_F, \mathcal{H}_N \otimes \mathcal{H}_F, D_f \otimes 1 + \gamma \otimes D_F, J \otimes J_F, \gamma \otimes \gamma_F\right).
$$
In this picture $D_F$ is usually considered fixed and controls physical aspects of the gauge fields, while $D_f$ will vary and describe the metric structure of the fuzzy space $M_f$.
The smooth limit of the fuzzy Yang-Mills-Higgs triple is given by $N \to \infty$, while the size of the gauge sector $n$ remains fixed.
As in Equation \ref{general fuzzy Dirac} we can write $D_f = \sum \gamma^I \otimes \{K_I, \cdot\}_{e_I}$, the amplification $D_f \otimes 1$ can then be written as $\sum \gamma^I \otimes \{K_I \otimes 1_n, \cdot\}_{e_I}$ acting on $V_{p, q} \otimes M_N(\mathbb{C}) \otimes M_n(\mathbb{C})$.

The gauge theory enters this framework in the guise of inner fluctuations.
These inner fluctuations \cite{Connes Inner Fluctuations} (see also chapter 6 of \cite{VS}) arise from Morita self-equivalences and are parametrised by the self-adjoint \textit{Connes one-forms}, for a spectral triple $(\mathcal{A}, \mathcal{H}, D, J, \gamma)$ these one-forms are given by
$$
  \Omega^1_D(A) = \left\{ \sum a_i[D, b_i] \,\middle|\, a_i, b_i \in A\right\} \subset \End(\mathcal{H})
$$
where the sum is finite.
For a self-adjoint $\omega \in \Omega^1_D(A)$, the \textit{fluctuated Dirac operator} is given by $D_\omega = D + \omega + J \omega J^{-1}$.

For the fuzzy Yang-Mills-Higgs triple we can split the fluctuations into a fuzzy and a finite part, by $a[D, b] = a[D_f \otimes 1, b] + a[\gamma \otimes D_F, b]$ for $a, b \in M_N(\mathbb{C}) \otimes M_n(\mathbb{C})$.
The fuzzy part of this, $a[D_f \otimes 1, b]$, can be parametrized by $\Omega^1_{D_f}(M_N(\mathbb{C})) \otimes M_n(\mathbb{C})$, while the finite part can, independently, be parametrized by $M_N(\mathbb{C}) \otimes \Omega^1_{D_F}(M_{n}(\mathbb{C}))$.
The effect of the fuzzy part of the fluctuation, for $s \neq 1, 5$, on $D_f \otimes 1$ is to replace $K_I \otimes 1_n$ by $K_I \otimes 1_n + T_I$, with $T_I \in \Omega^1_{K_I}(M_N(\mathbb{C})) \otimes M_n(\mathbb{C})$ of the appropriate (skew-)adjointness.
The finite part of the fluctuation does not affect the fuzzy section, since it carries the action of $\gamma$ on the $V_{p, q}$ factor of $\mathcal{H}_N$.
Instead, the finite part of the fluctuation together with $D_F$ itself is gathered into one term $\Phi = 1_N \otimes D_F + \{\phi, \cdot \}_{\epsilon''}$ where $\epsilon''$ depends on $s$ and $\phi \in M_N(\mathbb{C}) \otimes \Omega^1_{D_F}(M_n(\mathbb{C}))$. $\Phi$ is suggestively called the Higgs potential.

To motivate the term Yang-Mills-Higgs spectral triple we will specialize to the 4-dimensional Riemannian case which has signature $(p, q) = (0, 4)$.
In this case the fuzzy Dirac operator can be written
$$
  D_f = \sum_\mu \gamma^\mu \otimes [L_\mu, \cdot] + \gamma^{\hat{\mu}} \otimes \{X_\mu, \cdot\}.
$$
Here $\gamma^{\hat{\mu}}$ is the increasingly ordered product of the gamma matrices for $V_{0,4}$ except $\gamma^\mu$.
The $(0, 4)$ fuzzy Dirac operator shows remarkable similarity to the Dirac operator on a commutative manifold, with $[L_\mu, \cdot]$ taking the place of $\partial_\mu$ and $\{X_\mu, \cdot \}$ taking the place of, appropriately symmetrized, Christoffel symbols.
This leads us to say the fuzzy space is flat if all $X_\mu$ vanish.

The term Yang-Mills is further motivated by considering the gauge group \cite{Connes Inner Fluctuations, QFTNCG, VS} associated to a spectral triple $(A, H, D, J, \gamma)$,
$$
  G(A, J) = \{uJuJ^{-1} \,|\, u \in U(A)\}.
$$
For a Yang-Mills fuzzy geometry, this gauge group is $PU(N) \times PU(n)$ acting in the adjoint representations on $M_N(\mathbb{C}) \otimes M_n(\mathbb{C})$.
The $PU(N)$ factor corresponds to symmetries of the base fuzzy geometry, while $PU(n)$ acts on the finite ``vector bundle'' part as a Yang-Mills gauge group.

When basing a Yang-Mills-Higgs fuzzy geometry on such a flat $(0, 4)$ fuzzy space the self-adjoint one-forms, and thus the inner fluctuations, can be parametrized as $\sum_\mu \gamma^\mu \otimes A_\mu $ with $A_\mu \in \Omega^1_{L_\mu}(M_N(\mathbb{C})) \otimes M_n(\mathbb{C})$ skew-adjoint.
The effect of such a fluctuation on $D_f$ is by replacing $L_\mu$ by $L_\mu + A_\mu$, making the action of inner fluctuations analogous to having a connection on the ``vectorbundle'' $F$.
In this setting we define the field strength of a Dirac operator by $F_{\mu\nu} = \left[[L_\mu + A_\mu, \cdot], [L_\nu + A_\nu, \cdot] \right]$, mimicking the regular commutative definition of the field strength of a connection.

Considering the quartic potential $S(D) = g \tr D^2 + \tr D^4$ we have \cite{Sanchez3}, for a fluctuated $(0, 4)$ Dirac operator,
$$
  S(D) = -2\tr F_{\mu\nu}F^{\mu\nu} + 4\tr(g \theta + \theta^2) + 4\tr(g\Phi^2 + \Phi^4) - 8\tr([L_\mu + A_\mu, \Phi][L^\mu + A^\mu, \Phi])
$$
where $\theta = \eta^{\mu \nu}[L_\mu + A_\mu, \cdot ] \circ [L_\nu + A_\nu, \cdot]$. Here $\theta$ is analogous to a Laplace operator and the trace is taken as operator on $M_N(\mathbb{C}) \otimes M_n(\mathbb{C})$.
Each of these four tracial terms has an interpretation in terms of commutative Yang-Mills-Higgs theory:
\begin{itemize}
  \item $\tr F_{\mu\nu}F^{\mu\nu}$ is analogous to the classical Yang-Mills action,
  \item $\tr g\theta + \theta^2$ contains geometric information similar to a Laplace operator through a type of heat-kernel expansion \cite{Spectral estimators},
  \item $\tr g\Phi^2 + \Phi^4$ represents, for appropriate values of $g$, the Higgs potential,
  \item $\tr ([L_\mu + A_\mu, \Phi][L^\mu + A^\mu, \Phi])$ is the coupling between the Yang-Mills connection and the Higgs field.
\end{itemize}

Hence we can include a Yang-Mills-Higgs action in a Dirac ensemble by taking the product with a finite spectral triple and considering inner fluctuations.
The resulting Dirac ensemble has a space of Dirac operators $\mathcal{D}$ parametrized by, in the flat signature $(0,4)$ case, the four skew-adjoint matrices $L_\mu$ as well as the inner fluctuations $A_\mu \in \Omega^1_{L_\mu}(M_N(\mathbb{C}))$ and  $\phi \in M_N(\mathbb{C}) \otimes \Omega^1_{D_F}(M_n(\mathbb{C}))$.
The physical $D_F$ is considered fixed.
It should be noted that the $A_\mu$ are not necessarily independent, as they originate from a single one-form $\omega \in \Omega^1_{D_f}(M_N(\mathbb{C}))$ further complicating any analytical investigation of these models.

Investigating Yang-Mills-Higgs ensembles numerically or for a lower dimensional fuzzy spectral triple would be very interesting.
Another open direction is the addition of a fermionic term to the action in the Dirac ensemble of the form $\langle D \psi, \psi \rangle$ where $\psi \in \mathcal{H}_N \otimes \mathcal{H}_F$ plays the role of a fermion field and would be assigned a probability distribution together with the metric and gauge fields represented by $D$ and its inner fluctuations.

\section{Spectral statistics and phase transitions}\label{Phase transitions}
In this section  we discuss  spectral statistics and phase transitions that have been studied for Dirac ensembles so far.  In Section 2.1 we discuss general properties of the spectra of random Dirac ensembles. Next, in Section 2.2 we look at the phase transition in the large $N$ limit of spectral density functions of Dirac ensembles and present a new result.  In Section 2.3 we discuss the manifold-like behavior of spectra of Dirac ensembles at various phase transition points and attempts to make this idea more concrete.  Finally, in Section 2.4 we show that in certain Dirac ensembles one can recover the critical exponents and partition functions of minimal models from Liouville quantum gravity. We recommend that readers unfamiliar with the spectral density functions and genus expansions of random matrix ensembles review Appendices \ref{convergent integrals} and \ref{formal matrix models}.

\subsection{Spectral statistics of Dirac ensembles}

The Dirac ensembles of type $(1,0)$ and $(0,1)$ discussed in Sections \ref{One dimensional Dirac} can be analyzed as bi-tracial matrix models using various standard random matrix techniques that are applicable to that setting. Consider a type $(1,0)$ or $(0,1)$ Dirac ensemble with a partition function of the form
\begin{equation*}
	Z = \int_{\mathcal{D}} e^{- \frac{t_{2}}{4}\tr D^{2} + \sum_{j=3}^{d}\frac{t_{2j}}{4j}\tr D^{2j}}dD.
\end{equation*}
The  spectra of $H$ and $L$ appearing in the associated matrix model can be computed, as we will see in subsequent sections. However, we are not just interested in the spectrum of $H$ and $L$ but also in the spectrum of $D$.  It was first conjectured in \cite{Barrett2015} and later proven in \cite{Second paper} that if the limiting eigenvalue distribution, $\rho(x)$, of the associated random matrix ensemble exists then the limiting  eigenvalue density function of $D$, $\rho_{D}$(x), is given by the integral convolution of the random matrix spectral density function with itself i.e.
\begin{equation*}
	\rho_{D}(x) = \int_{\mathbb{R}}\rho(x-t)\rho(t)dt.
\end{equation*}

The relationship between spectral densities is far from clear for higher signature Dirac ensembles, even for the two dimensional ones from Section \ref{two dim Dirac}. Moreover, the associated matrix models of these Dirac ensembles are multi-trace and multi-matrix, of which little is known \cite{Eynard combin}.

In the large $N$ limit, there is a  universality to the spectral density function of Dirac ensembles for any signature when the potential is Gaussian. In \cite{Barrett2015} a Gaussian potential 
\begin{equation*}
	S(D) = \tr D^{2},
\end{equation*} 
is investigated. When looking at the $(1,0)$ Dirac ensemble, the associated matrix potential becomes
$2N\tr H^{2} + 2 \tr H \tr H$, where $H$ is a Hermitian matrix. For  the $(0,1)$ Dirac ensemble, this potential becomes
$2N\tr L^{2} + 2 \tr L \tr L $ where $L$ is skew-Hermitian.
For these ensembles the numerics show that the distribution of the eigenvalues of  $H$  and $L$ resembles Wigner's semicircular distribution as $N$ increases. This suggests that the multi-trace term has little to no impact as $N$ gets larger.  
Dirac ensembles of signatures $(2,0)$, $(1,1)$ and $(0,2)$ were also studied for small matrix size, where some of the above-mentioned results also apply. 
Furthermore, the eigenvalue density function of the Dirac operator was conjectured to be the integral convolution of Wigner's semicircular law with itself. 

This was then proven in \cite{Second paper}. It was shown that in the large $N$ limit the eigenvalue density function for $D$ is universal, in the sense that for any signature $(p,q)$ the limit is the same given the right scaling.
Consider the partition function of the form
\begin{equation*}
	Z = \int_{\mathcal{D}}e^{-\frac{1}{2k}\tr D^{2}}dD
\end{equation*}
where $k$ is the dimension of $V_{p,q}$. Then the density function of $D$, for any signature $(p,q)$, is in the large $N$ limit given by
\begin{equation*}
	\rho_{D}(x) = \int_{\mathbb{R}}\rho_{W}(x-t)\rho_{W}(t)dt,
\end{equation*}
where
\begin{equation*}
	\rho_{W}(x) = \frac{1}{2\pi}\sqrt{4-x^{2}}_{[-2,2]} 
\end{equation*}
is Wigner's Semicircular Distribution \cite{Second paper}. 

\begin{figure}[H]\label{Wigner}
	\includegraphics[width=8cm]{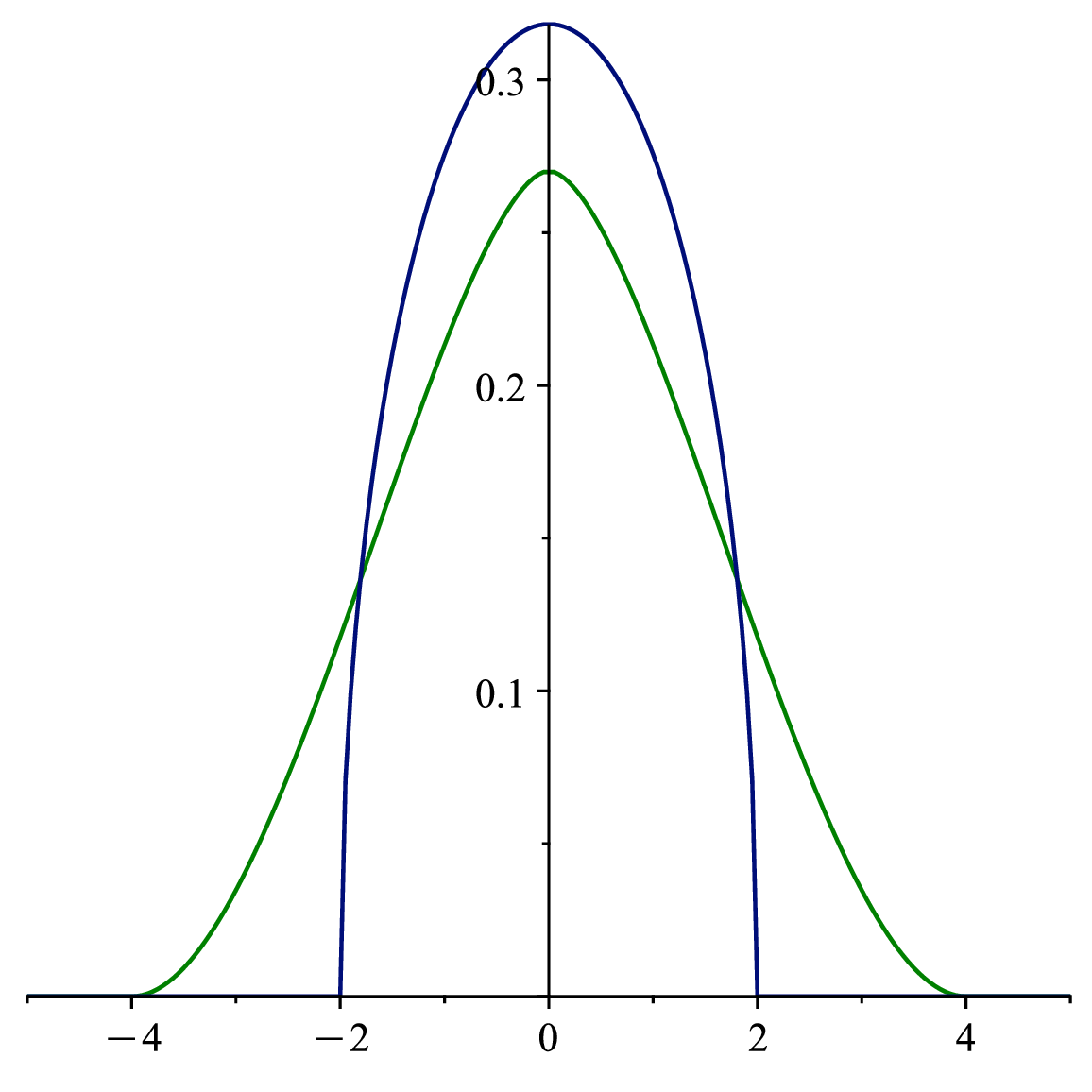}
	\centering
	\caption{The Wigner semicircular distribution in blue compared to the Wigner  convolution distribution in green.}
\end{figure}

\subsection{Spectral phase transitions} \label{spectral phase transition}
In this subsection we will discuss results related to \textit{spectral phase transitions}, which we define as a configuration of coupling constants where the number of connected components of the support of the large $N$  eigenvalue density function of the random matrix ensemble changes. Like their random matrix model cousins, Dirac ensembles also exhibit phase transitions. This was first investigated numerically using Markov chain Monte Carlo methods in \cite{Barrett2016} and then  in \cite{glaser}. It was later proven rigorously in \cite{First paper} that spectral phase transitions indeed occur in quartic type $(1,0)$ and $(0,1)$ Dirac ensembles. In this subsection we recall this quartic phase transition from \cite{First paper} and we also present a new result proving a phase transition for a sextic  Dirac ensemble.  It is an interesting problem to analytically investigate phase transitions for Dirac ensembles of any type $(p, q)$ and a potential of any order. We believe that these phase transitions  do indeed exist, but we  don't have a proof at hand. The existing results use what is known as the  Coulomb gas method, the rigorous foundations of which are fully developed in \cite{Deift}.  We give a brief summary of this technique in Appendix \ref{convergent integrals}.

For Gaussian potentials, unsurprisingly, no phase transition exists. This motivated the study of a Landau-Ginzburg type quartic potential as studied in \cite{Barrett2016}. This potential is of the form 
\begin{equation*}
	S(D) = g_{2} \tr D^{2} + \tr D^{4},
\end{equation*} 
and was studied numerically for different small signatures $(p,q)$ for a matrix size of ten in \cite{Barrett2016}. It was discovered that, for many signatures, the spectrum of the Dirac operator displayed a single-cut distribution for certain values of $g_{2}$ which transitioned into a double-cut regime for different $g_2$. This suggests the existence of a spectral phase transition. Furthermore, it was noted that near the phase transition the spectrum of $D$  asymptotically behaves like the Dirac operator on a two  dimensional manifold i.e. $\rho_{D} (\lambda) \sim C_{D} |\lambda|$, as $\lambda$ goes to infinity, where $C_{D}$ is a constant.

Computing these eigenvalue density functions explicitly even in the large $N$ limit is very difficult. Dirac ensembles of dimension higher than one with a potential more complicated than the Gaussian are multi-trace multi-matrix models about which little is known. Usual methods such as orthogonal polynomials cannot be applied because of the multi-trace terms, Weyl's integration formula cannot be applied because of the lack of unitary invariance, and the loop equations are  too complicated for Topological Recursion. Furthermore, it is not of the form of a Harish-Chandra integral, and is too complicated for a characteristic expansion. For a review of these techniques see \cite{Random Matrices}. 

However, for Dirac ensembles of signature $(1,0)$ or $(0,1)$, several options are available. The multi-trace terms still prevents the use of orthogonal polynomials, but the Coulomb gas technique can be applied. This was done in \cite{First paper} where the type $(1,0)$ quartic model  
\begin{equation*}
	Z = \int_{\mathcal{D}}e^{-g \tr D^2 - \tr D^4 }dD = \int_{\mathcal{H}_{N}}e^{-2N(g \tr H^2 +   \tr H^4) -2g (\tr H)^2 - 8 \,\tr H \tr H^3 - 6(\tr H^2)^2 }dH
\end{equation*}
was studied. We will recall those results here as an example. In our paper \cite{First paper}, the phase transition location is off due to a missing scalar factor. In this paper we will give the new correct value that we have been able to derive analytically and have also verified with our own Monte Carlo simulations. Furthermore, the relationship between the coupling constant and the support is given by slightly different equations in both the single cut and double cut solutions.  It was shown in \cite{First paper} that the quartic $(1,0)$ and $(0,1)$ ensembles have the same behavior in the large $N$ limit. Using the Coulomb gas method as explained in Appendix \ref{convergent integrals}, we obtained the following explicit formula for the limiting eigenvalue density function of $H$: for $g > -4\sqrt{2}$,
\begin{align*}
\rho(x)&=  \frac{1}{\pi}(-4\gamma^{2} +\frac{1}{2\gamma^{2}} + 4x^{2})\sqrt{4\gamma^{2}-x^{2}}_{[-2\gamma,2\gamma]},
\end{align*}
where  the support $[-2\gamma,2\gamma]$ can be found as a function of $g$ as the root of

\begin{align*} 
192 \gamma^8 + 48 \gamma^4 + 4 g \gamma^2 -1 = 0.
\end{align*}
 
\begin{figure}[H]
	\centering
	\begin{subfigure}{0.5\textwidth}
		\centering
		\includegraphics[width=.9\textwidth]{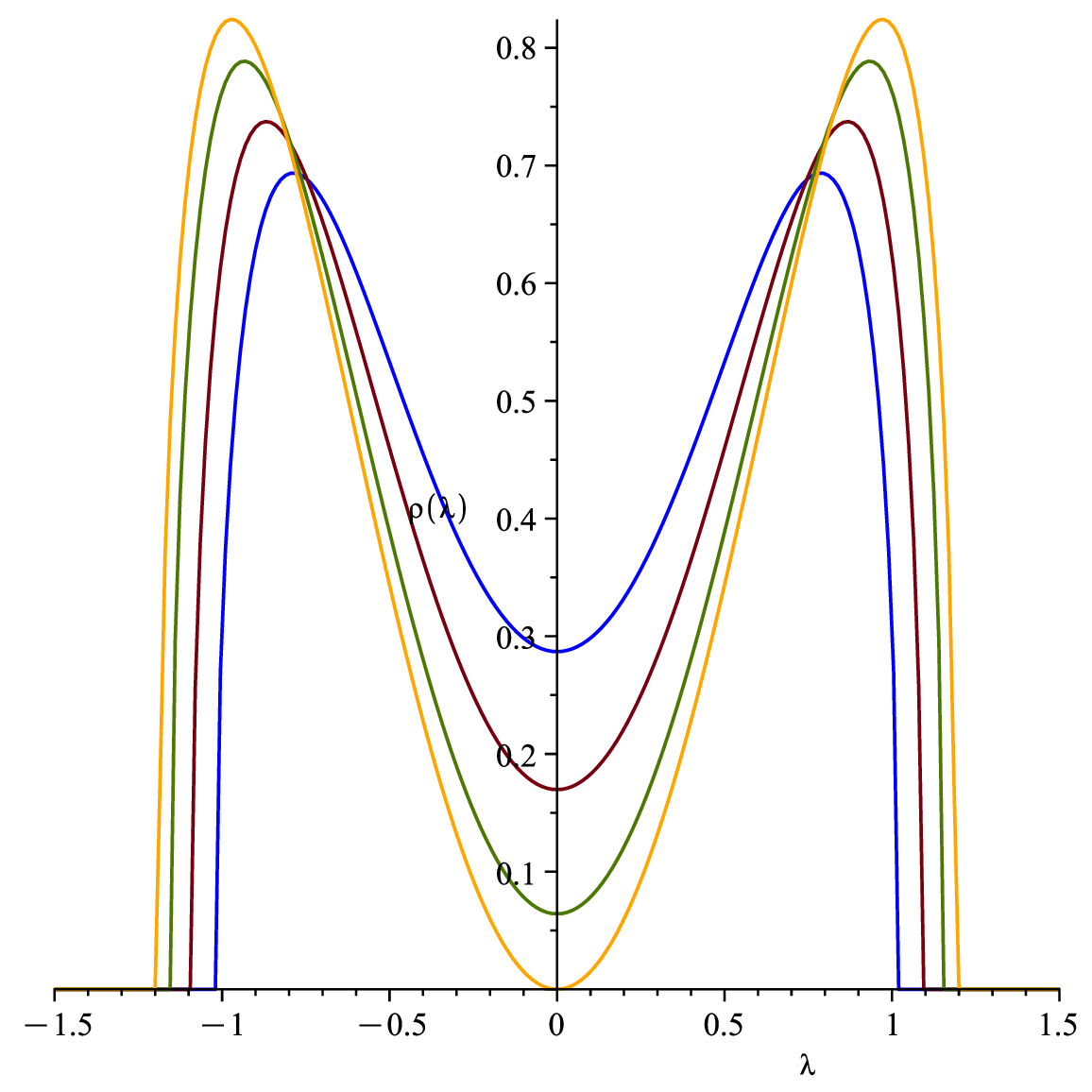}
	\end{subfigure}%
	\begin{subfigure}{.5\textwidth}
		\centering
			\includegraphics[width=.9\textwidth]{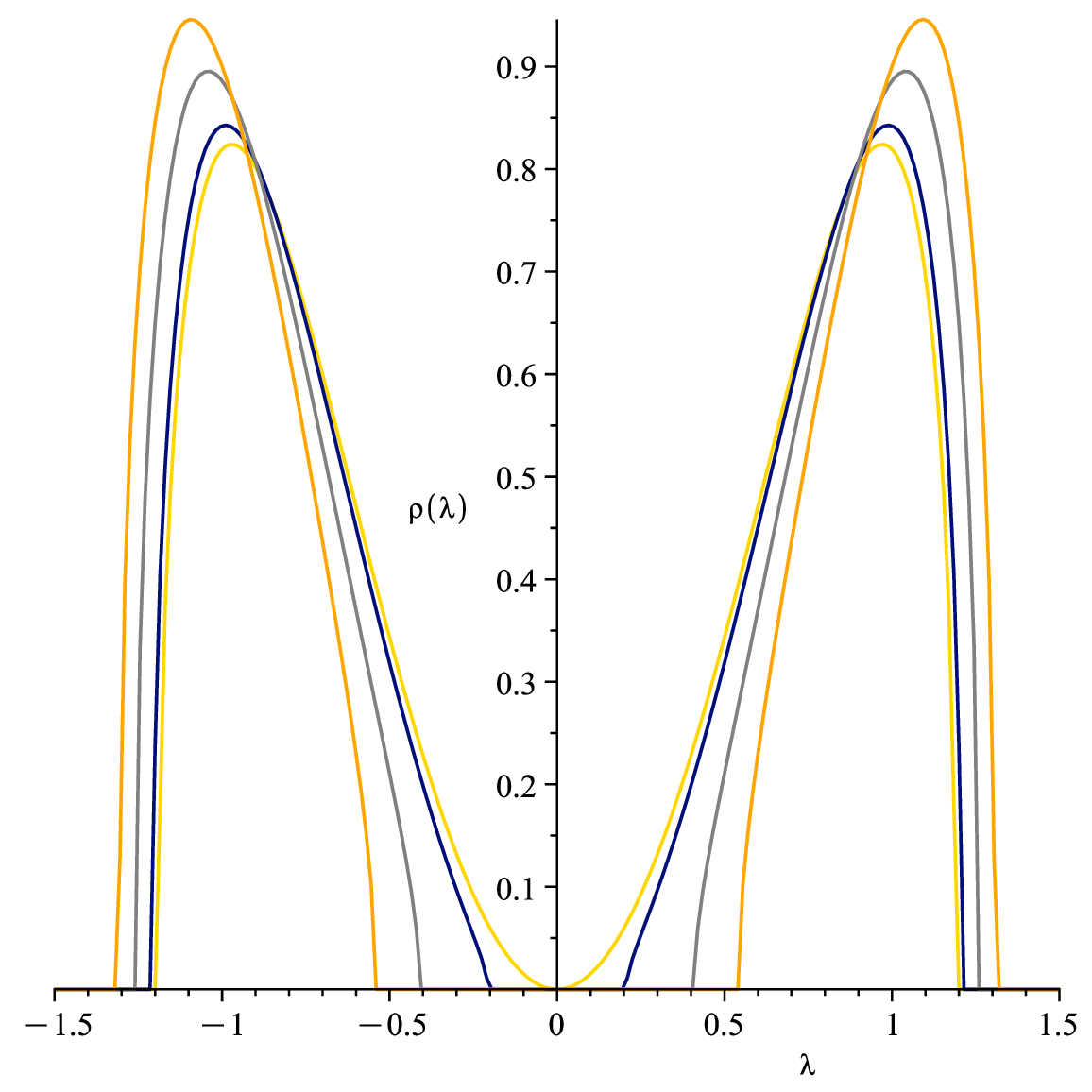}
	\end{subfigure}
	\caption{The eigenvalue density functions  for  type $(1, 0)$ and $(0, 1)$ quartic matrix models in the large $N$ limit.  The colors in the above figures  correspond to different values of $g$ as follows: blue is $g= -3$, red is $g=-4$, green is $g=-5$, yellow is $g = - 4 \sqrt{2}$, navy is $g=-6$, gray is $g=-7$, and orange is $g=-8$.} 
	\label{DOS1}
\end{figure}
When $g =-4\sqrt{2}$ the spectral phase transition occurs. For $g < -4\sqrt{2}$ the new limiting eigenvalue density function was found to be

\begin{align*}
\begin{split}
\rho(x) = \frac{2}{\pi}|x| \sqrt{(x^{2}-a^{2})(b^{2}-x^{2})}_{[-a,-b]\cup [b,a]},
\end{split}
\end{align*}
where the support $[-a,-b]\cup [b,a]$ can be found in terms of $g$ via the equations

\begin{equation*}\label{a in terms of g}
a^{2}= -\frac{1}{8}g +\frac{1}{\sqrt{2}},
\end{equation*}
and 
\begin{equation*}\label{b in terms of g}
b^{2}= -\frac{1}{8}g -\frac{1}{\sqrt{2}}.
\end{equation*}
These  results are plotted in Figure \ref{DOS1}.

\begin{figure}[H]
	\centering
	\begin{subfigure}{0.5\textwidth}
		\centering
		\includegraphics[width=0.9\linewidth]{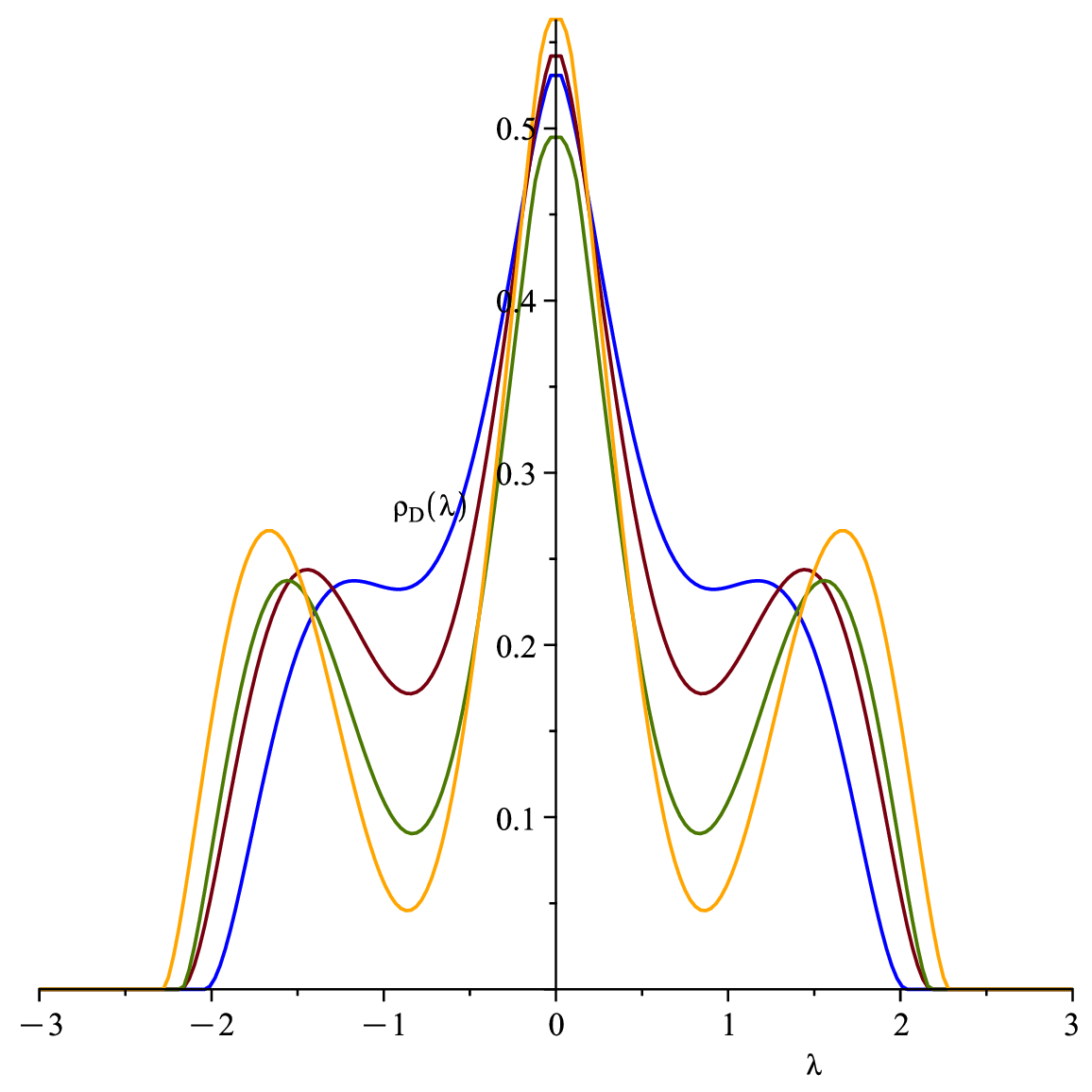}
		\label{fig:sub1}
	\end{subfigure}%
	\begin{subfigure}{0.5\textwidth}
		\centering
		\includegraphics[width=0.9\linewidth]{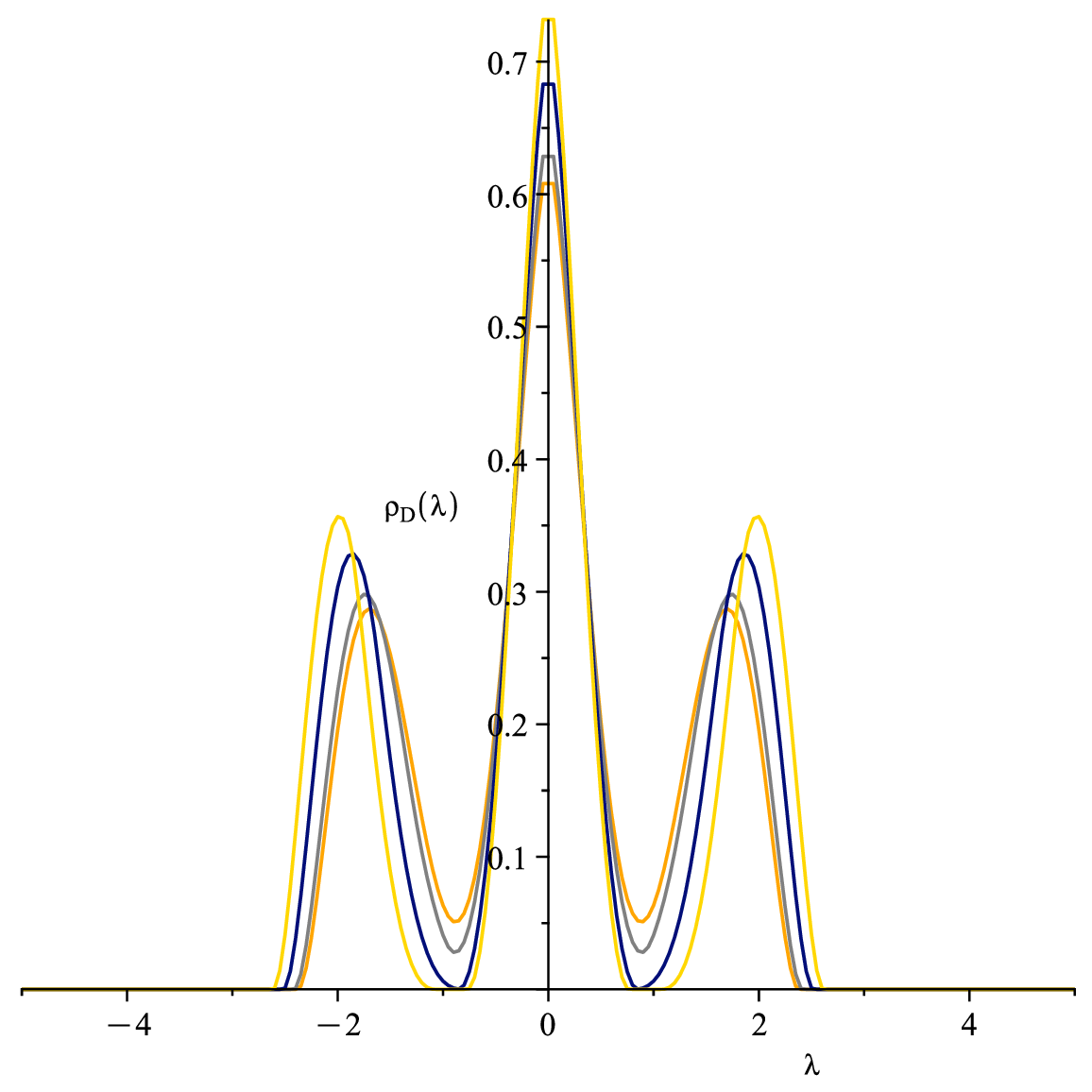}
		\label{fig:sub2}
	\end{subfigure}
	\caption{The eigenvalue density functions of the $(1, 0)$ and $(0, 1)$ quartic Dirac ensembles in the large $N$ limit. The colors in the above figures  correspond to different values of $g$ as follows: blue is $g= -3$, red is $g=-4$, green is $g=-5$, yellow is $g = - 4 \sqrt{2}$, navy is $g=-6$, gray is $g=-7$, and cyan is $g=-8$.}
	\label{DOS2}
\end{figure}

As discussed in the previous section, the eigenvalue density function of the Dirac operator in the large $N$ limit is the convolution of the density functions of $H$. See Figure \ref{DOS2}.

We note that the  techniques presented in \cite{First paper} in fact apply to any even potential Dirac ensembles of signature $(1,0)$ or $(0,1)$ and any convergent bi-tracial single matrix model. For example, let us consider the following type $(1,0)$ sextic model  
\begin{equation*}
	Z = \int_{\mathcal{D}}e^{-\frac{g}{2} \tr D^2 - \frac{1}{6}\tr D^6 }dD. 
\end{equation*}
Once again employing the Coulomb gas method results in the following explicit formula for the limiting eigenvalue density function: 

\begin{align*}
	\rho(x)&=\frac {-40\,{{\gamma}}^{12}+20\,{{ \gamma}}^{10}{x}^{2}+10\,{ 
					\gamma}^{8}{x}^{4}+50\,{{\gamma}}^{10}-50\,{{\gamma}}^{8}{x}^{2}+24\,{
				{\gamma}}^{6}-12\,{{\gamma}}^{4}{x}^{2}-{{\gamma}}^{2}{x}^{4}-1}{20
			\,{{\gamma}}^{8}\pi-2\,\pi\,{{\gamma}}^{2}}\sqrt {4\,{{\gamma}}^{2}
			-{x}^{2}}_{[-2\gamma,2\gamma]}
\end{align*}
where  the support $[-2\gamma,2\gamma]$ can be found as a function of $g$ by (numerically) solving the equation

\begin{align*} \label{1,0 poly}
	 1=10\,\frac {{{\gamma}}^{2} \left( 60\,{{\gamma}}^{16}-130\,{{\gamma}}^{14}-26\,{{\gamma}}^{10}-22\,{{\gamma}}^{8}+{{\gamma}}^{6}{
	 			g}-5\,{{\gamma}}^{4}-{g}/10 \right) }{10\,{{\gamma}}^{6}-1
	 }
	 .
\end{align*}

Alternatively one can use the Lagrange inversion formula to find a perturbative expansion of $\gamma$ in terms of $g$.  
This model also exhibits at least one spectral phase transition and in general the plots are very similar to the quartic model. However, it is the belief of the authors that it might exhibit another phase transition, but further analysis is required. For details of this analysis, which is identical to the formal case, see the sextic example in \cite{HKP 2}.

\begin{figure}[H]
	\centering
		
	\begin{subfigure}{0.5\textwidth}
		\centering
		\includegraphics[width=.8\textwidth]{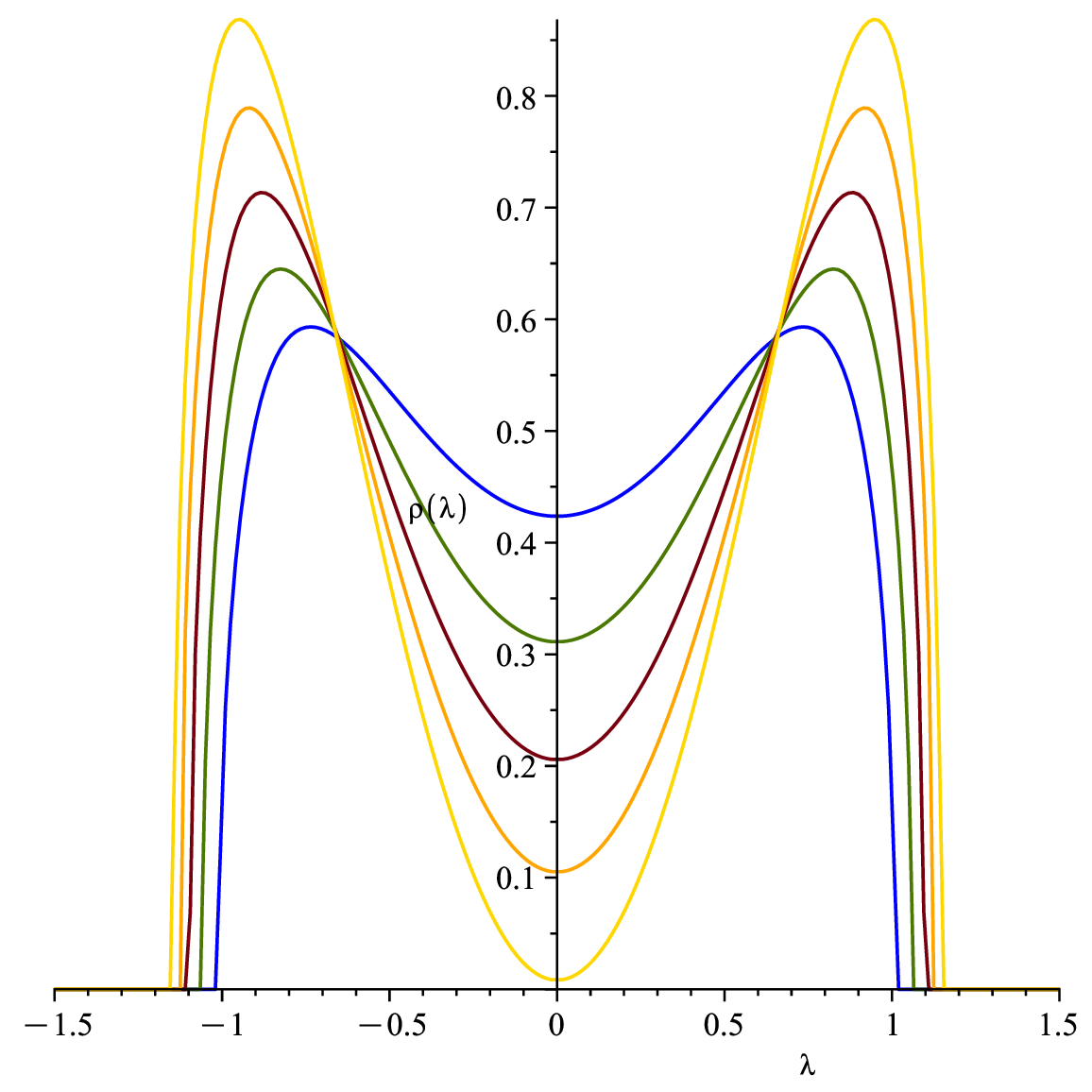}
	\end{subfigure}%
	\begin{subfigure}{.5\textwidth}
		\centering
		\includegraphics[width=.8\textwidth]{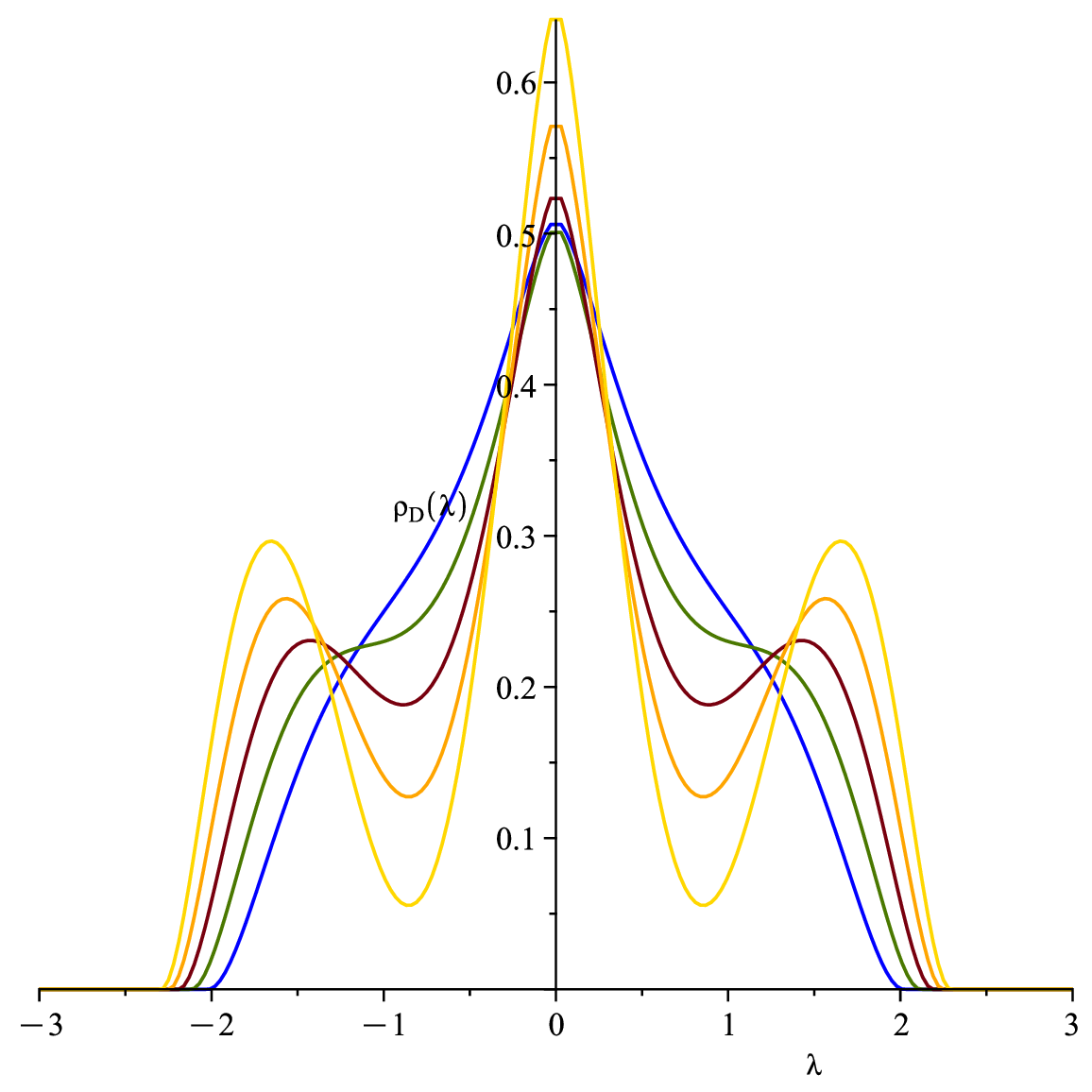}
	\end{subfigure}
	\caption{The eigenvalue density functions  for  type $(1, 0)$ sextic matrix model (left) and for the corresponding Dirac ensemble (right) in the large $N$ limit. The colors in the above figures  correspond to different values of $g$ as follows: blue is $g = -1.5$, green is $g =-4 $, red is $g= -6.5$, orange is $g=-9$, and yellow is $g=-11.5$. }
	\label{DOS3}
\end{figure}

\subsection{Spectral geometry of fuzzy spaces}
In this subsection we shall very briefly give an overview of some of the results obtained in \cite{Spectral estimators}. A simple, yet interesting, model of a discrete  and finite noncommutative  spin Riemannian manifold  is a fuzzy spectral triple as we defined in Section 1. Given the important role that spectral geometry has played in global analysis, geometry, and topology of manifolds, as well as in noncommutative geometry, it is natural to ask to what extent its ideas can be extended to fuzzy spectral triples. 

However, many of the results from spectral geometry are based on asymptotics of the spectrum $(\lambda_n)_{n \in \mathbb{N}}$ of the Dirac operator $D$ (ordered such that $|\lambda_{n+1}| \geq |\lambda_n|$).
Since the spectrum of the Dirac operator of a finite spectral triple is finite there is an obvious problem in directly generalizing such asymptotic results from spectral geometry to fuzzy spaces. For example, the first major result of spectral geometry is the celebrated Weyl's asymptotic law, according to which the volume and dimension of a compact $d$-dimensional Riemanian manifold can be recovered from its Dirac spectrum:
$$ \text{Vol}(M) = \lim_{n \rightarrow \infty} \frac{n\,(4 \pi)^{d/2} \,\Gamma(1+d/2)}{k\, |\lambda_{n}|^{d}},$$
where $k$ is the rank of the spinor bundle of $M$ and $\Gamma(s) = \int_{0}^{\infty}e^{-t}\frac{t^{s}}{t}dt$ is the Gamma function. 

In \cite{Spectral estimators} the authors define, and numerically investigate, quantities computed from the spectrum that recover the usual asymptotic properties of dimension and volume for infinite spectra, but are also applicable to finite spectra.
One such quantity is the spectral variance which is defined as 
\begin{equation*}
	v_{s}(t) = 2 t^2 \left(\frac{\sum_{i} \lambda_{i}^{4}e^{-\lambda_{i}^{2}t}}{\sum_{i} e^{-\lambda_{i}^{2}t}}- \left(\frac{\sum_{i} \lambda_{i}^{2}e^{-\lambda_{i}^{2}t}}{\sum_{i} e^{-\lambda_{i}^{2}t}}\right)^{2}\right).
\end{equation*}
If computed for the spectrum of a manifold one obtains $\lim_{t \to 0} v_s(t) = \dim(M)$, which follows from Weyl's law.
This quantity is also sensible for finite spectra, and gives the expected result of two for the fuzzy sphere and tori for appropriate values of $t$ \cite{Spectral estimators}.
The presence of the parameter $t$ should be thought of as an energy scale, with $t$ small corresponding to high energies and small wavelengths and $t$ large corresponding to low energies and large wavelengths.
For fuzzy spaces $\lim_{t\to 0} v_s(t) = 0$, as fuzzy spaces have a minimum wavelength, or Planck length, built in.

Another way to access asymptotic information in the spectrum $(\lambda_n)$ is by the spectral zeta function
$$\zeta_{D} (s)=\sum_{n=1}^{\infty} \frac{1}{\lambda_{n}^{2s}}, $$
which turns out to be useful even in the finite case.
It is well-known that Weyl's Law is equivalent to the fact that the volume of $M$ can be expressed in terms of the residue of the spectral zeta function $\zeta_{D} (s)$ at its top pole $s = d/2$:
$$ \text{Vol}(M) = \frac{(4\pi )^{d/2}}{k}\text{Res}_{s=d/2}( \zeta_{D}(s)\Gamma(s)).$$

One can use the spectral zeta function to define a notion of distance between fuzzy spectral triples and even between fuzzy spectral triples and manifolds.
One possibility for a notion of distance between metric spaces is the Gromov-Hausdorff distance \cite{Gromov}, but in \cite{Cornel} the authors define a distance notion more useful in this situation, based on the spectra of Riemannian manifolds. Let $D_{1}$ and $D_{2}$ be two Dirac operators with $\zeta_{D_1}(s)$ and $\zeta_{D_2}(s)$ their spectral zeta functions. Then the distance between geometries is defined to be 
\begin{equation}\label{spectral distance}
	\sigma (D_{1},D_{2}) = \sup_{\gamma \leq s \leq \gamma +1}\left| \log \left(\frac{\zeta_{D_1}(s)}{\zeta_{D_2}(s)}\right)\right|
\end{equation}
for some interval $[\gamma,\gamma +1]$ where all poles lie below $\gamma$. For Dirac operators on compact spin  manifolds it was found in \cite{Cornel} that this is indeed a metric, in particular $\sigma(D_{1},D_{2}) = 0$ if and only if the spectra are the same.

In \cite{Spectral estimators} this idea is adapted to define a distance between (random) fuzzy spectral triples.
For example for each fuzzy sphere with matrix size $N$ we have a Dirac operator of size $N$. This spectrum is the same as the Dirac operator on the spin bundle of the 2-sphere tensored with $\mathbb{C}^{2}$ but with a cut-off. As $N$ goes to infinity the spectral zeta function of the fuzzy sphere converges to the spectral zeta function of the sphere uniformly on the interval $[\gamma, \gamma +1]$ for any $\gamma >1$. Thus,  $\sigma(D_{N},D_{S^{2}}) $ goes to zero as $N$ goes to infinity. Note that when considering truncated spectra, pointwise convergence of zeta functions is not a sufficient condition for $\sigma(D_{1},D_{2}) = 0$, uniform convergence is needed. For more on fuzzy spaces and truncated spectral triples see \cite{truncated 1, truncated 2}.

\begin{figure}[H]
	\centering
	\includegraphics[width=0.7\textwidth]{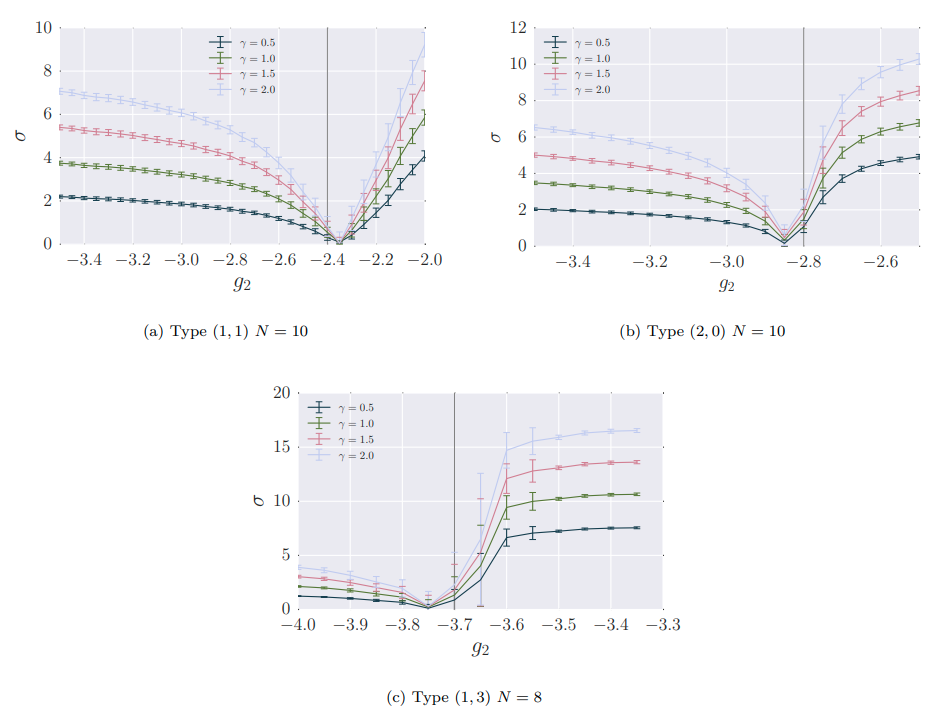}
	\caption{The spectral distance metric (\ref{spectral distance}) is plotted between various type $(p,q)$ quartic Dirac ensembles of various matrix sizes and the fuzzy sphere \cite{Spectral estimators}. Note that the spectral distance goes to zero near where phase transitions (denoted by a vertical line) were found to exist in \cite{Barrett2016,glaser}.}
	\label{spectral est}
\end{figure}

One of the most remarkable results in \cite{Spectral estimators} is that, when using the spectral distance to compare random spectra sampled from each of the quartic Dirac ensembles of types $(1,1)$, $(2,0)$, and $(1,3)$ to the fuzzy sphere for various values of the coupling constant, near the spectral phase transition the spectral distance $\sigma$ tends to zero. See Figure \ref{spectral est}.
The authors of \cite{Spectral estimators} found further numerical evidence that, near spectral phase transitions, Dirac ensembles display manifold-like behavior, but we are still far from proving such a conjecture rigorously.

\subsection{Liouville quantum gravity }\label{Liouville quantum }
It has long been known that random matrix theory has connections to two dimensional quantum gravity. These range from the famous Kontsevich model \cite{Witten, Kontsevich}, to the newly discovered connections to Jackiw-Teitelboim (JT) gravity \cite{JT 1, JT 2, JT 3, JT 4}. Of particular interest however, is the connection to conformal field theory. Heuristically, physicists knew in the 80's and 90's from the asymptotics of convergent matrix integrals (found using orthogonal polynomials) that certain matrix ensembles have critical exponents corresponding to models in conformal field theories coupled to gravity \cite{LQG 1, LQG 2}. 

The idea is that matrix models count maps which can be thought of as discretized Riemann surfaces. If the coupling constants of the models are tweaked such that the number of polygons that form these maps goes to infinity, one would in essence be counting Riemann surfaces, which are also counted in conformal field theories in two dimensions.  Later in  \cite{Eynard LQG},  this idea was made precise with formal matrix models. These formal models often have the same asymptotics as their convergent matrix model counterparts (if such a counterpart exists). We will briefly describe the idea behind these critical points here.
 
Suppose a formal random matrix model's partition function $Z$ has the following genus expansion
\begin{equation*}
 	\log Z = \sum_{g \geq 0} N^{2-2g} \,F_{g},
\end{equation*}
where the $F_{g}$ are the generating functions of certain types of maps (specified by the model) with no boundaries \cite{Eynard2018}, see Appendix \ref{formal matrix models} for details. Random matrix models often have critical points where the coupling constants of the model are such that the number of polygons in the maps goes to infinity. In such cases it is proven that the $F_{g}$ have asymptotic expansions around these points, which have critical exponents corresponding to a minimal model in conformal field theory \cite{Ds multi 2, Eynard LQG}. Additionally, these matrix model asymptotics satisfy a partial differential equation that is also satisfied by the corresponding minimal model. For example, the quartic Hermitian matrix model corresponds to the $(3,2)$ minimal model, also referred to as pure gravity, and its partition function satisfies Painlev\'e I.
 
How  this relates to Dirac ensembles is not as obvious at a first glance.  However, for several models examined in \cite{HKP 2} we prove that single Hermitian matrix models are hidden in Dirac ensembles. In particular the cubic, quartic, and sextic Dirac ensembles of type $(1,0)$ contain the cubic, quartic, and sextic Hermitian matrix models, respectively, in their phase space. This is nontrivial because in a Dirac ensemble coupling constants are not attached to specific bi-tracial terms. Instead many single and bi-tracial terms share the same coupling constants,  so one cannot easily turn off all bi-tracial terms by setting certain coupling constants to zero.

Consider for example the quartic Dirac ensemble from Section \ref{One dimensional Dirac}. As alluded to, one can fine-tune its coupling constants such that we recover a quartic Hermitian matrix model in the large $N$ limit. For a full proof see \cite{HKP 2}. 

\begin{figure}[H]
	\centering
	\includegraphics[width=0.5\textwidth]{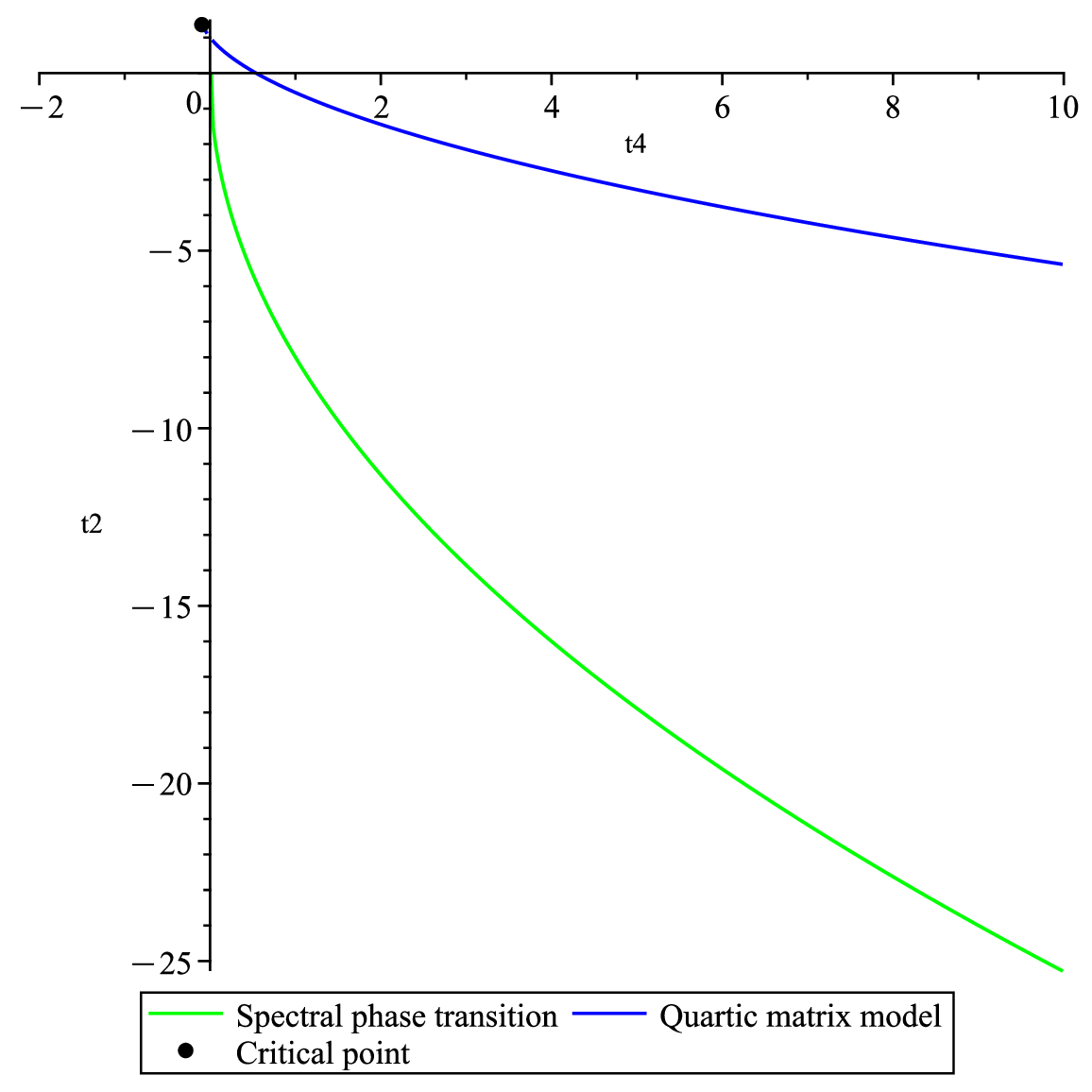}
	\caption{The phase diagram of the quartic Dirac ensemble \cite{HKP 2}. The y-axis is $t_{2}$ and the x-axis is $t_{4}$.}
	\label{fig:quartic diagram}
\end{figure}

The  single trace quartic model 
\begin{equation}\label{quartic}
	\int_{\mathcal{H}_{N}}e^{-\frac{N}{2}\tr H^{2} -  \frac{t_{4}}{ 4}N\tr H^{4}}dH
\end{equation}
has a critical point  at $t_{4} =- 1/12$ which we will refer to as $t_{c}$ \cite{Eynard2018}. The singular parts of the $F_{g}$ are algebraic for all genus $g$, except for $g = 1$,  in which case it is logarithmic. In particular they are
\begin{equation*}
	\text{sing}(F_{g}) = C_{g}\, (t_{4}-t_{c})^{5(1-g)/2}
\end{equation*}
and
\begin{equation*}
	\text{sing}(F_{1}) = C_{1} \,\log(t_{4}-t_{c})
\end{equation*}
for $g \not =1$. These expansions are often referred to as a ``double scaling limit''. We may define the generating series
\begin{equation*}
u(y) = \sum_{g= 0}^{\infty} C_{g} y^{5(1-g)/2}.
\end{equation*}
Then   $u''(y)$ satisfies the Painlev\'e I  equation,  i.e.
\begin{equation*}
y = (u''(y))^{2} - \frac{1}{3} u^{(4)}(y).
\end{equation*}
We know that the critical point of the quartic Hermitian model is $t_4 = -1/12$ \cite{Eynard2018},  and that the quartic Dirac ensemble 
\begin{equation}\label{Dirac quartic}
Z = \int_{\mathcal{D}}e^{-\frac{t_{2}}{4} \tr D^2 -\frac{t_{4}}{8} \tr D^4 }dD
\end{equation}
contains it i.e. for a certain choice of coupling constants you recover the quartic Hermitian model. In particular, this happens when $t_{2} = 4/3$ and $t_{4} = -1/12$ \cite{HKP 2}. Figure \ref{fig:quartic diagram} allows one to visualize several phenomena of the model in a phase diagram.

In this diagram we have the curve where a spectral phase transition occurs, this is a generalization of the phenomenon in Section \ref{spectral phase transition} with two coupling constants. The equation of this curve is
\begin{equation*}
t_2 =-8 \sqrt{t_{4}}.
\end{equation*}
Below the critical (green) curve  the spectrum of the Dirac ensemble is in a 3-cut phase and above it 1-cut, as seen in Section \ref{spectral phase transition}.

The diagram also shows for what values of $(t_{2},t_{4})$ the Dirac ensemble in equation (\ref{Dirac quartic}) has the quartic Hermitian matrix model, equation (\ref{quartic}), as the associated matrix model if the coupling constants lie on the blue curve with equation:
\begin{equation*}
t_2=-\frac{(1 + 12 t_4)^{3/2} - 4 - 144 t_4 + (36 t_4 + 3) \sqrt{1 + 12 t_4}}{72 t_4}.
\end{equation*}
These equations are derived in \cite{HKP 2} by solving the loop equations. Note that  these curves do not intersect because the quartic Hermitian matrix model as stated in equation (\ref{quartic}) does not have a spectral phase transition. The reason is that there is no coupling constant in front the $\tr H^{2}$, which is required to reach the phase transition.

Hermitian multi-matrix models  are also associated  with  a much wider class of minimal models than single matrix Hermitian matrix models \cite{DS multi 1, Ds multi 2}. Thus it would be interesting to investigate numerically if higher signature Dirac ensembles  are also associated to minimal models. The question also remains unproven as to which $(1,0)$ or $(0,1)$ Dirac ensembles correspond to minimal models. Additionally, there might also be other critical points in Dirac ensembles besides those mentioned or other interesting critical phenomena.

The authors would like to emphasize that none of the connections between conformal field theory and random matrix theory  mentioned here are new. The point is rather to lend credit to a noncommutative theory of quantum gravity where integrating over finite Dirac operators in place of metrics can be used to recover other toy models of quantum gravity.

\section{Bootstrapping the loop equations}\label{boot}

{\it Bootstrapping} was introduced in elementary particle physics as part of the S-matrix program by Geoffrey Chew in the early 1960's. The idea was to use any consistency conditions available to compute various correlation functions of interest  and especially to formulate a theory of strong interactions. The mantra was  ``particles pull themselves up by their own bootstraps''. But, after an initial success,  the idea stalled in producing  viable new results and predictions.   Meanwhile a  competing theory, the standard model of elementary particles, was created based on the theory of   quarks   and  gauge theory,  which  could  indeed successfully account for experimental data. As a result,  bootstrap methods   were nearly  forgotten for a long time. 
In recent years,  however,  there has been a revival of the bootstrap idea  mostly  thanks to  the  success of the conformal bootstrap program  by Rattazzi,  Rychkov, and collaborators in 2008  in understanding phase transition and critical phenomena in dimensions bigger than two \cite{RRTV}. In two dimensions, the conformal bootstrap was demonstrated to work in 1983 by  Belavin,  Polyakov and  Zamolodchikov \cite{BPZ}.

In the context of random matrix theory and related fields, bootstrapping recently emerged in several works, first by Anderson and  Kruczenski in the context of lattice gauge theory \cite{Anderson}. In a   random matrix setting  bootstrapping  was first used  by Lin \cite{bootstraps},  then  in our paper
\cite{HKP}, and also by Kazakov and Zheng \cite{Kazakov}. Bootstrapping  has  also recently been applied  to  matrix quantum mechanics as well \cite{Matrix QM, numerical bootstraps,  Berenstein}. In the following section we present a new example that deals with a cubic Dirac ensemble as well as recalling our results from \cite{HKP}. To readers unfamiliar the with Schwinger-Dyson equations in the context of random matrix theory we recommend reviewing Appendix \ref{SDE's and TR}.

 In another interesting new development, bootstrapping is now used in computing the  spectrum of Einstein and hyperbolic manifolds in \cite{Bon1, Bon2}. The eigenvalues of the Laplace-Beltrami operator, as well as the integrals of their eigenfunctions, satisfy certain positivity conditions that imply bounds on both quantities. One wonders whether an extension of these ideas to some classes of spectral triples is possible.

\subsection{\texorpdfstring{The cubic type $(1,0)$ Dirac ensemble}{The cubic type (1, 0) Dirac ensemble}}
We will start with a brief overview of how bootstrapping works for Dirac ensembles.
The large $N$ limits of higher moments of random matrix models satisfy an infinite  system of nonlinear equations,  which was first derived by Migdal \cite{Migdal}. These so-called {\it loop equations} are consequences of Schwinger-Dyson equations and the factorization property of moments at large $N$ limits.
In general the loop equations are not restrictive enough to fully determine the moments. However, as we shall explain later in this section, one can bring to bear some positivity  constraints on moments to further restrict the set of possible solutions to the loop equations. The process of further narrowing the search space by adding certain extra positivity constraints is called {\it bootstrapping}.   Further positivity constraints are obtained from the fact that our matrix models  stem  from  Dirac operators of spectral triples. This  extra positivity  is quite useful and is missing in   general matrix models.  By narrowing down the search space one can sometimes recover the values of the initial moments. From there the loop equations can be used, in theory, to find any moment.

We will now give a novel example of the bootstrap method by applying it to a cubic Dirac ensemble. Using the bootstrap technique we are able to find a relationship between the coupling constant of the model, the first moment and from there higher moments. Let us consider a  type \((1,0)\) cubic Dirac ensemble with the partition function
\begin{equation*}
	Z = \int_{\mathcal{D}} e^{-\frac{1}{4}\tr D^2 - \frac{g}{6} \tr D^3 }dD.  
\end{equation*}
This integral is obviously not convergent but understood perturbatively, i.e. as a formal matrix model. The associated matrix model is a bi-tracial Hermitian single-matrix model with the potential
\begin{equation*}
	\tilde{S}(H) = \frac{1}{2} \left(N \tr H^2 +  \left(\tr H\right)^2 \right) + \frac{g}{3}\left( N \tr H^3 + 3 \tr H^2 \tr H \right).
\end{equation*}

The loop equations of the model are then given by: 

\begin{align}\label{cubic loop eqns}
	\sum_{k = 0}^{\ell-1} m_k m_{\ell-k-1} =  m_{\ell+1} + m_1 m_{\ell} + g \left( m_{\ell+2} + 2 m_1 m_{\ell+1} + m_2 m_{\ell} \right) \quad \text{for }\ell \in \mathbb{N} \cup \{0 \}.
\end{align}
In particular we get the following loop equation for $\ell = 0$: 

\begin{align*}
	g \left( m_2 + m_1^2 \right) + m_1 = 0.
\end{align*}
It can be shown that by having the first moment $m_1$, we can recursively calculate higher moments using the loop equations. We refer to such a situation as the dimension of the search space being one. If the loop equations required $n$ moments or higher moments to determine all other (higher) moments, we would say the search space has dimension $n$. 

The existence of an eigenvalue density function $\rho(x)$, which is a probability density function, gives us constraints on moments in the following way. Take a real polynomial  $f(x) = \sum_{j=1}^{k} c_j x^j$. Then by the non-negativity of the integral we have
$$ \int_{\mathbb{R}} f(x)^{2} \rho(x) dx = \sum_{i,j=1}^{k} \int_{\mathbb{R}}c_{i}c_{j}x^{i+j}\rho (x)dx = \sum_{i,j=1}^{k}c_{i}c_{j}m_{i+j} \geq 0,$$ 
for all real values of the $c_{i}$.
This shows that the quadratic form $\sum_{i,j=1}^{k}c_{i}c_{j}m_{i+j}$ is positive semi-definite. Since this holds for all $k = 1,2,...$ and it can be expressed nicely in terms of the positive semi-definiteness of the Hankel matrix of moments

\begin{align*}
\begin{bmatrix}
m_0   &  m_1 &  m_2 & m_3 & \cdots \\ 
m_1 &  m_2 &  m_3 & m_4 & \cdots \\
m_2 &  m_3 &  m_4 & m_5 & \cdots \\
m_3 &  m_4 &  m_5 & m_6 & \cdots \\
\vdots & \vdots &  \vdots & \vdots & \ddots
\end{bmatrix}\geq 0.
\end{align*}
As a side remark we should add that Hamburger's Theorem says that positivity of this matrix is a necessary and sufficient condition for a sequence of real numbers $m_{0},m_{1},m_{2},...$ to be the moments of a probability distribution, see page 145 of \cite{Reed and Simon}.

\begin{figure}[H]
	\centering
	\includegraphics[width=0.7\textwidth]{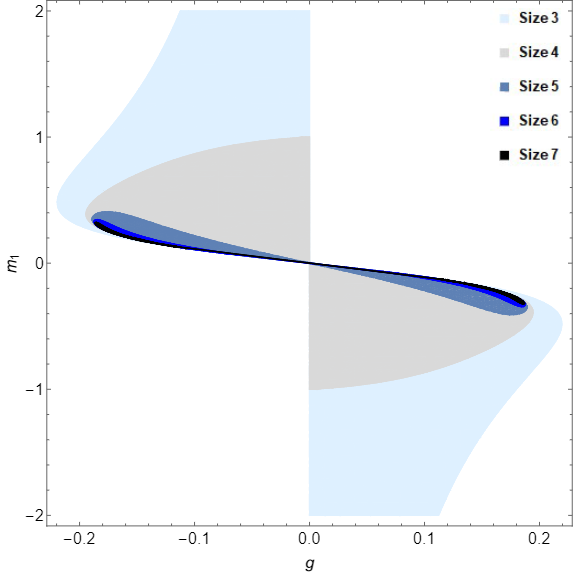}
	\caption{The constraints on the relation between $g$ and $m_1$ for the (1,0) cubic model found by bootstrapping. Each colour corresponds to a different number of constraints derived from positivity of principal minors. The solution space narrows as the number of constraints increases. Notice that in this example increasing the number of constraints shows that there exists a nonlinear relationship between $g$ and $m_{1}$. Additionally the analytic solution from \cite{HKP 2} is plotted for comparison.}
	\label{cubic bootstrap}
\end{figure}

The positive semi-definiteness of the Hankel matrix tells us, in particular, that every leading principal minor is greater than or equal to zero. This gives us countably many inequalities, or constraints, involving the moments. Combining this observation with the fact that in this particular model every moment can be written in terms of $m_{1}$, thanks to the structure of the loop equations (\ref{cubic loop eqns}), we obtain an infinite number of nonlinear  constraints on $m_{1}$. Using semidefinite programming to find the region satisfied by these constraints gives us Figure \ref{cubic bootstrap}.

It is worth noting that positivity constraints can be applied to both the Hankel matrix of moments of the matrix ensemble and the Hankel matrix of moments of the Dirac ensemble. As mentioned earlier the moments of the Dirac operator are defined as 
\begin{equation*}
	d_{\ell} = 	\lim_{N \rightarrow \infty}\left\langle \frac{1}{N^2}\tr D^{\ell}\right\rangle=\lim_{N \rightarrow \infty} \frac{1}{N^2}\, \frac{1}{Z}\int_{\mathcal{D}}\tr D^{\ell} e^{-\frac{1}{4}\tr D^2 - \frac{g}{6} \tr D^3 } dD,
\end{equation*}
so by the same argument as before we have
\begin{align*}
	\begin{bmatrix}
		1   &  d_1 &  d_2 & d_3 & \cdots \\ 
		d_1 &  d_2 &  d_3 & d_4 & \cdots \\
		d_2 &  d_3 &  d_4 & d_5 & \cdots \\
		d_3 &  d_4 &  d_5 & d_6 & \cdots \\
		\vdots & \vdots &  \vdots & \vdots & \ddots
	\end{bmatrix} \geq 0.
\end{align*}
These additional constraints are a particular advantage that Dirac ensembles have over ordinary matrix ensembles when bootstrapping.

We emphasize that this model can be solved analytically \cite{HKP 2}, but is presented here as an insightful example of the bootstrap technique. 

\subsection{\texorpdfstring{The quartic type $(2,0)$ Dirac ensemble}{The quartic type (2, 0) Dirac ensemble}}
Bootstrapping can also be applied successfully to ensembles that are, to the best of our knowledge, unsolvable. 
To illustrate this let us consider the quartic action for a type $(2,0)$ ensemble, which appears in \cite{Barrett2016,glaser,Spectral estimators},
\begin{equation*}
Z=	\int_{\mathcal{D}} e^{-g\tr D^{2} - \tr D^{4}}dD, 
\end{equation*}
where the associated matrix potential is given by equation (\ref{quartic (2,0)}). We will summarize the results of \cite{HKP} to show the effectiveness of bootstrapping for this model. Note that the action of this model is symmetric under the transformations
\begin{align*}
	D &\rightarrow -D,\\ 
	H_{1} &\rightarrow -H_{1},\\
		H_{2} &\rightarrow -H_{2},\\
		&\text{and}\\
			H_{1} &\leftrightarrow H_{2}.
\end{align*} 
This greatly simplifies the loop equations.  In particular all   the odd moments and odd higher moments are zero i.e. any moment of a word in $H_{1}$ and $H_{2}$ containing either an odd number of $H_1 $ or $H_2$ is zero. After considering these symmetries the following terms of the potential are the ones that contribute to the loop equations in the large $N$ limit
\begin{align*}
& g( 4N\tr H_1^2 + 4N\tr H_2^2) + 4N \tr H_1^4 + 4N \tr H_2^4  \\ 
&+16 N\tr H_1^2H_2^2 -8N \tr H_1H_2H_1H_2 
+ 12  (\tr H_1^2)^2 \\
 &+ 12  (\tr H_2^2)^2 + 8 \tr H_1^2 \tr H_2^2.
\end{align*}
We use the following loop equations for bootstrapping:
\begin{align*}\label{loop equations (2,0)}
\sum_{k=0}^{\ell-1} m_k m_{\ell-k-1} = (8g+64m_2) m_{\ell+1} + 16m_{\ell+3}-16m_{\ell,1,1,1}  +32 m_{\ell+1 , 2 }
\end{align*}
in the large $N$ limit where we denote mixed moments as
\begin{equation*}
m_{a,b,c,d} = \lim_{N \rightarrow \infty}\frac{1}{N} \langle \tr H_1^a H_2^bH_1^cH_2^d \rangle. 
\end{equation*}
 Positivity constraints for mixed moments can be derived but require a slightly more general setting.

\begin{figure}[H]
	\centering
	
	\begin{subfigure}{0.5\textwidth}
		\centering
		\includegraphics[width=.8\textwidth]{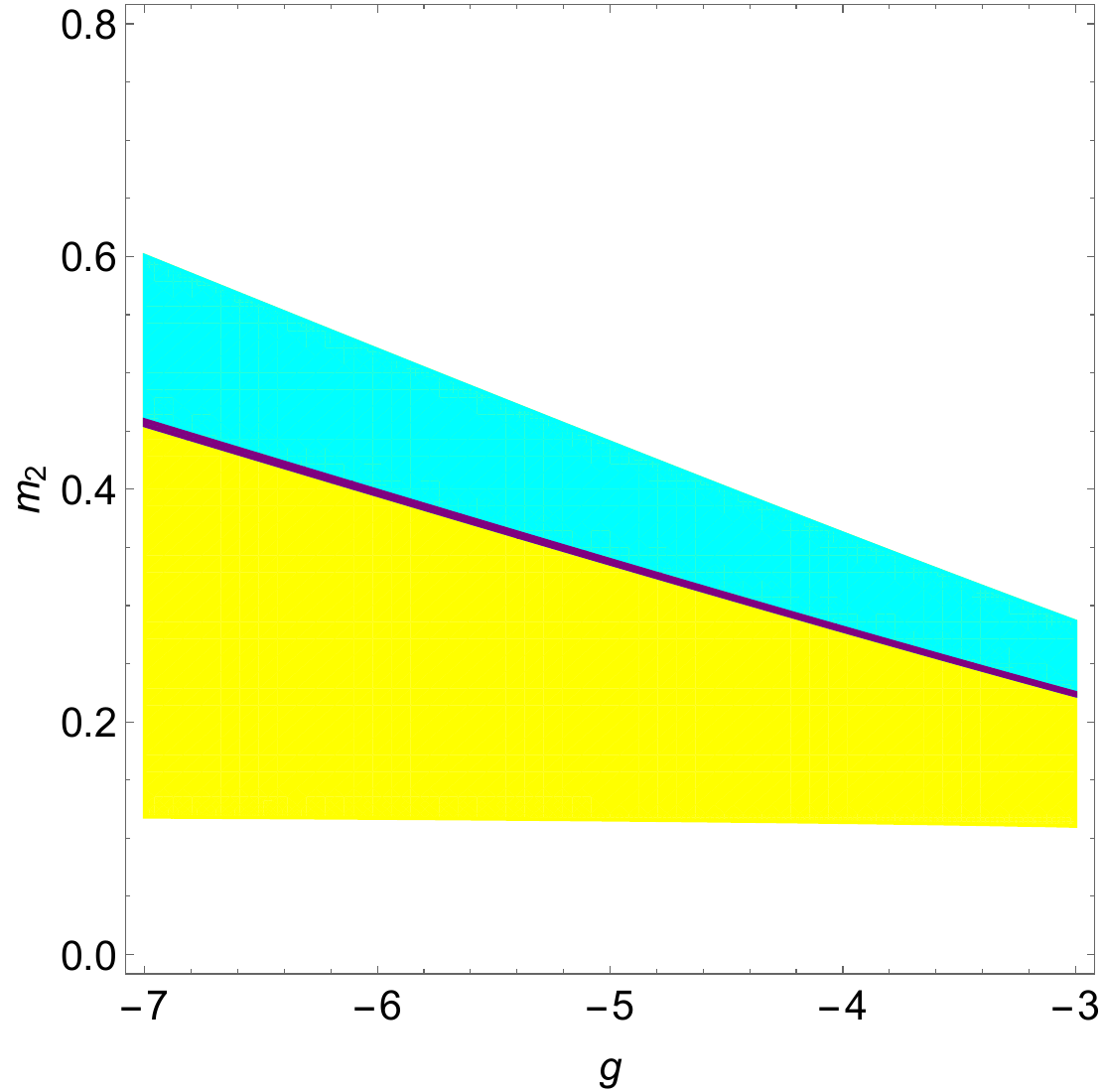}
	\end{subfigure}%
	\begin{subfigure}{.5\textwidth}
		\centering
		\includegraphics[width=.8\textwidth]{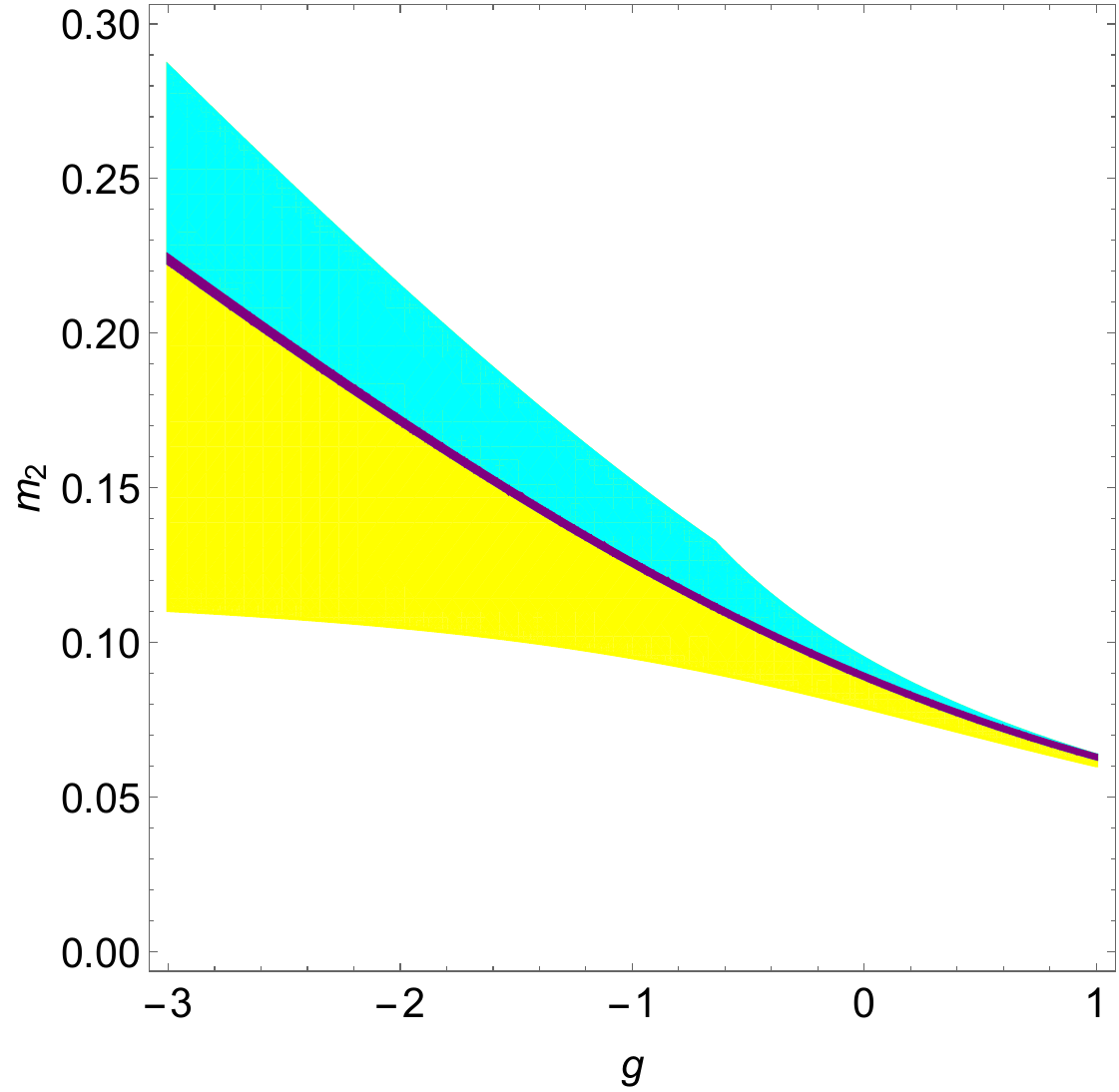}
	\end{subfigure}
	\caption{The search space region for the (2,0) quartic Dirac ensemble m. Note that near the phase transition found in \cite{Barrett2016} the  relationship between the coupling constant $g$ and $m_{2}$ appears to change from non-linear to linear \cite{HKP}.}
	\label{fig:quartic 2 0 search space}
\end{figure}

 Unlike the approach in \cite{bootstraps}, we considered the moments of all words. By doing this we were able to prove in \cite{HKP} that the search space for this model has dimension one. The results are displayed in Figure \ref{fig:quartic 2 0 search space}.
One interesting feature is that the relation between $m_{2}$ and $g$ appears to be linear for values of $g$ below the phase transition  \cite{glaser}. This is also  what was observed analytically in the type $(1,0)$ quartic in \cite{First paper}.

By generating all the loop equations using \textit{Maple} we found the following remarkable formulas for moments in terms of $g$ and the second moment $m_{2}$:

\begin{equation*}
m_{4} = -\frac{1}{8}gm_2 + \frac{1}{64},
\end{equation*}
\begin{equation*}
m_{2,2} = -\frac{1}{8}gm_2 -m_{2}^2 + \frac{1}{64},
\end{equation*}
\begin{equation*}
m_{1,1,1,1} = \frac{gm_{2}}{8} + 2 m_{2}^2 - \frac{1}{64},
\end{equation*}
\begin{equation*}
	m_{6} = {\frac {{g}^{2}m_2}{64}}-{\frac {g}{512}}-{\frac {g{m_{2}}^{2}}{8}}+{\frac {
			3\,m_{2}}{64}}-{\frac {5\,{m_{2}}^{3}}{4}},
\end{equation*}
\begin{equation*}
	m_{4,2} = {\frac {{g}^{2}m_{2}}{64}}+{\frac {g{m_{2}}^{2}}{8}}-{\frac {g}{512}}-{\frac {
			{m_{2}}^{3}}{4}}+{\frac {m_{2}}{64}},
\end{equation*}
\begin{equation*}
	m_{3,1,1,1} = -{\frac {{g}^{2}m_{2}}{64}}-{\frac {3\,g{m_{2}}^{2}}{8}}-{\frac {7\,{m_{2}}^{3}}{4
	}}+{\frac {g}{512}}+{\frac {m_{2}}{64}},
\end{equation*}
\begin{equation*}
	m_{2,1,2,1} = {\frac {{g}^{2}m_{2}}{64}}+{\frac {3\,g{m_{2}}^{2}}{8}}-{\frac {g}{512}}+{
		\frac {11\,{m_{2}}^{3}}{4}}-{\frac {m_{2}}{64}},
\end{equation*}

\begin{equation*}
m_{8}=-{\frac {gm_{2}}{64}}+{\frac {{m_{2}}^{4}}{4}}+{\frac {{g}^{2}}{4096}}+{\frac 
		{{m_{2}}^{2}}{256}}+{\frac{3}{4096}}-{\frac {{g}^{3}m_{2}}{512}}+{\frac {3\,{g
			}^{2}{m_{2}}^{2}}{64}}+{\frac {g{m_{2}}^{3}}{2}}.
\end{equation*}

 Note that the trace powers and therefore moments of this Dirac ensemble do not have a clear formula. These trace powers were studied closely in \cite{Sanchez}. With the above formulas and those borrowed from \cite{Sanchez} we have
 \begin{equation*}
 	d_{2} = 8\, m_{2},
 \end{equation*}
 \begin{equation*}
 	d_{4} = -4\,gm_2+{\frac{1}{2}},
 \end{equation*}
\begin{equation*}
d_{6} = -160\,{m_2}^{3}-16\,g{m_2}^{2}+6\,m_2+2\,{g}^{2}m_2-{\frac {1}{4}}g.
\end{equation*}

\section{Summary and outlook}
In this paper we gave an overview of the recent efforts to utilize random matrix theory techniques to give insight into toy models of Euclidean quantum gravity suggested by noncommutative geometry and initially proposed in \cite{Barrett2016}. We saw that type $(1,0)$ or $(0,1)$ Dirac ensembles can be analyzed analytically using the Coulomb gas technique \cite{First paper} and with the Blobbed Topological Recursion of stuffed maps \cite{AK, Second paper}.  However, for ensembles with dimension higher than one no known analytic techniques of random matrix theory seem to apply. Instead they may be examined at finite matrix size using Monte Carlo simulations \cite{Barrett2016,glaser} or at large matrix size using bootstrap techniques \cite{bootstraps}. Most recently it was discovered that certain Dirac ensembles are dual to minimal models in conformal field theory \cite{HKP 2}. It is worth investigating if this is true for more types and potentials. Additionally one wonders if connections to other theories of quantum gravity are possible, such as the recent connection found between random matrix theory, Topological Recursion, and JT gravity \cite{JT 1,JT 2,JT 3,JT 4}. 

One naturally wants a coupling  of these models with fermions and gauge fields. In noncommutative geometry there is a  finite spectral triple $F$ of the standard model of elementary particles \cite{QFTNCG, VS}. In their   work,   Chamseddine, Connes and Marcolli \cite{Neutrino Mixing}  consider spectral triples of the form $X \times F$, where $X$ is a Riemannian manifold  which  represents the gravitational sector.   Using the spectral action principle \cite{Spectral action} and heat kernel expansion they were able to obtain the Lagrangian of the standard model coupled with gravity. For a recent account and survey see \cite{Advances in NCG}.  Instead of a manifold,  we can take  a noncommutative space and a finite real spectral triple as  a good first approximation to that.  We should mention that the initial steps in this direction has already been taken in \cite{Gesteau}, and also in \cite{Sanchez3}. Especially in the latter work the kinematics of coupling  the gravitational field, in the context of finite spectral triples, with the Yang-Mills-Higgs  field of the standard model is worked out. What remains to be done is to choose a suitable potential $S$  to go in the path integral, to  calculate various quantities of interest, and also to study the large $N$ and double scaling limits of these quantities. We hope to come back to this project  in the  near future. 

An alternative approach to path integral quantization is the BV formalism. In the context of gauge theory on spectral triples this has been studied in \cite{BV Iseppi} and the BV formalism is applied directly to Dirac ensembles in \cite{BV nottingham}.

Finally let us suggest some open questions and problems in this line of research:
\begin{itemize}
		\item Investigate the limiting eigenvalue distribution of Dirac ensembles with more complicated potentials. This could be done numerically for any type or analytically for types $(1,0)$ and $(0,1)$ using the techniques outlined in \cite{First paper}.
	\item Do one dimensional Dirac ensembles have critical points other than those that appear when the coupling constants are tuned to become a single trace model \cite{HKP 2}?
	\item Do these new critical points have  double scaling limits of correlation functions that obey Blobbed Topological Recursion? This is seen in the single trace case \cite{Eynard2018}.
	\item Are there minimal models associated with higher dimensional Dirac ensembles? This could be investigated by looking at critical exponents using Monte Carlo simulations \cite{glaser} or perhaps using the functional renormalization group \cite{Sanchez2}.
	\item Can one find and make rigorous a connection between Dirac ensembles at phase transitions and the spectra of two dimensional manifolds, along the lines of \cite{Spectral estimators}?
	\item Is there a connection between Dirac ensembles and the recent work in noncommutative QFT \cite{Quartic K 1,Quartic K 2,Quart K 3,Quartic K 4}? One of the goals of this series of papers is  to prove that the correlation functions of the quartic Kontsevich model satisfy Blobbed Topological Recursion \cite{blobbed1}. As proved in \cite{AK}, certain Dirac ensembles also satisfy it. For a review see \cite{Quartic K review}.
	\item Apply the Batalin-Vilkovisky (BV) formalism of  \cite{BV nottingham} to  other Dirac ensembles.
	
	\item Investigate a possible relationship between Dirac ensembles and one-loop corrections to the spectral action  \cite{one loop}.
	
	\item Can we extend the Yang-Mills-Higgs theory to more general Dirac ensembles, coupling gravity with the standard model?  \cite{Sanchez3}.
	
	\item Investigate the consequences of adding a Fermionic term to the action of Dirac ensembles. 
	\end{itemize}

\begin{appendices}

	\section{Convergent unitarily invariant matrix ensembles}\label{convergent integrals}
	In general, a random matrix is a matrix valued random variable. In particular we are interested in random Hermitian and skew-Hermitian matrices.  The simplest example of interest is what is called the \textit{Gaussian Unitary Ensemble} (GUE). The joint distribution on the entries of a GUE matrix is given by
	\begin{equation*}
		\frac{1}{Z_{N}^{G}}e^{-\frac{N}{2}\tr H^{2}}dH,
	\end{equation*}
	where $Z_{N}^{G}$ is the normalization constant and $dH$ is the Lebesgue measure of the real $N^{2}$-dimensional vector space of $N\times N$ Hermitian matrices:
	\begin{equation*}
		dH = \prod_{i} d\text{Re} H_{ii} \prod_{i < j} d\text{Re} H_{ij}d \text{Im} H_{ij}.
	\end{equation*}
	
	The GUE gets its name from being invariant under unitary transformations on $H$. A much wider class of random matrix ensembles have this property and are referred to as (unitarily) invariant ensembles. Traditionally, the invariant ensembles of interest are given by measure a of the form
	
	\begin{equation*}
		\frac{1}{Z_{N}} e^{-\frac{N}{2}\tr H^{2} - \sum_{j= 3}^{d} \frac{N t_{j}}{j}\tr H^{j}}dH.
	\end{equation*}
	where $d$  is some even integer greater than three, and the $t_{j}$'s are some real coupling constants. 
	
	There are several quantities of interest when analyzing these ensembles. The partition function $Z_{N}$ can often be computed using orthogonal polynomials \cite{Deift}. However, when studying Dirac ensembles another type of invariant ensemble is considered, one where this technique is not applicable. Consider convergent matrix models of the form
	
	\begin{equation}\label{bi tracial potential}
		Z_{N}=\int_{\mathcal{H}_{N}} e^{-\frac{N}{2}\tr H^{2} -  \sum_{j = 3}^{d} \frac{N t_{j}}{j}\tr H^{j}-\sum_{i,j=1}^{d} \frac{ t_{i,j}}{i+j}\tr H^{i}\tr H^{j}}dH.
	\end{equation}
	Such models are called bi-tracial. 
	
	Another important quantity of interest are moments which we define as 
	\begin{equation*}
		\langle \frac{1}{N}\tr H^{\ell}\rangle   :=\frac{1}{N}\frac{1}{Z}\int_{\mathcal{H}_{N}} \tr H^{\ell} e^{-\frac{N}{2}\tr H^{2} -  \sum_{j=3}^{d} \frac{N t_{j}}{j}\tr H^{j}-\sum_{i,j=1}^{d} \frac{ t_{i,j}}{i+j}\tr H^{i}\tr H^{j}}dH
	\end{equation*}
for $\ell \geq 0$. In practice, given these moments one can find a unique  corresponding probability distribution with compact support. This quantity is often difficult to compute for finite $N$. However, calculations simplify if one considers computing the limit of these moments as $N$ goes to infinity, which we refer to as the large $N$ limit. This sequence of moments, in practice, also has a unique corresponding probability distribution, which is called the \textit{limiting eigenvalue distribution}.  In particular for convergent integrals of the form of (\ref{bi tracial potential}) it can be expressed as $d\mu(x) = \rho(x)dx$, where $\rho$ is a continuous function, referred to as the \textit{limiting eigenvalue density function}. For example, in the case of the GUE, the limiting eigenvalue distribution is the celebrated Wigner semicircular distribution.

Even though the direct computation of $Z_{N}$ for finite $N$ seems out of reach, one may still compute the limiting eigenvalue distribution in the large $N$ limit using Coulomb gas techniques, which we will now review. The first step to compute this distribution is to apply Weyl's integration formula \cite{Guionnet} to reduce the $N^{2}$-dimensional integral $Z_{N}$  to an integral over its $N$ eigenvalues: 
	\begin{align*}
		Z_{N}&=C_{N}\int_{\mathbb{R}^{N}} \exp\left\{-\frac{N}{2}\sum_{k=1}^{N} \lambda_{k}^{2} -  \sum_{j>2}^{d} \frac{N t_{j}}{j}\left(\sum_{k=1}^{N} \lambda_{k}^{j}\right)
		\right.\\
		&\qquad \qquad \qquad \qquad \left. -\sum_{i,j=1}^{d} \frac{ t_{i,j}}{i+j}\left(\sum_{k=1}^{N} \lambda_{k}^{i}\right)\left(\sum_{s=1}^{N} \lambda_{s}^{j}\right)\right\}\prod_{k<s}(\lambda_{k}-\lambda_{s})^{2} \prod_{k=1}^{N}d\lambda_{k}\\
		&=:C_{N}\int_{\mathbb{R}^{N}} \exp\left\{-\sum_{i,j=1}^{N} Q(\lambda_{i},\lambda_{j})+ 2 \sum_{i<j} \log |\lambda_{i}-\lambda_{j}|\right\} d\lambda_{k}
	\end{align*}
	for some constant $C_{N} $. Notice that the Jacobian from the change of variables gave us the square of the famous Vandermonde determinant in the integrand. 
	
	For certain potentials, the leading contribution of the integral is going to come from the set of eigenvalues that maximizes the integrand, we denote such a set $\{\lambda_{i}^{*}\}_{i=1}^{N}$. One can often show that such a point in $\mathbb{R}^{N}$ is unique, allowing us to apply Laplace's method. Furthermore, we may construct the normalized counting measure of eigenvalues:
	
	\begin{equation}\label{counting}
		\mu_{N} = \frac{1}{N}\sum_{i=1}^{N}\delta_{\lambda_{i}^{*}}.
	\end{equation}

	Using the results in chapter six of \cite{Deift} one can show that for  convergent integrals like (\ref{bi tracial potential}), the measure (\ref{counting}) converges in the vague topology to the limiting eigenvalue distribution of the ensemble. The limiting eigenvalue distribution of  (\ref{bi tracial potential}) can be found as the unique measure $\mu$ that minimizes the following functional:
	\begin{equation*}
		I(\mu) = \int_\mathbb{R}\int_\mathbb{R}\left(Q(x,y)-\log |x-y|  \right) d\mu(x) d\mu(y).
	\end{equation*} 
	With knowledge of $\mu$ one may compute the large $N$ moments of a random matrix ensemble as 
	\begin{equation*}
		\lim_{N \rightarrow \infty} \frac{1}{N}\langle \tr H^{\ell} \rangle =\int x^{\ell}d\mu(x). 
	\end{equation*} 
Additionally, though not as obvious how, one can also compute the free energy
\begin{equation*}
	\lim_{N \rightarrow \infty}\frac{1}{N^{2}}\ln Z_{N}
\end{equation*}
from knowledge of the spectral density $\rho$. For more details see chapter 6 of \cite{Deift}.
	
	Define the \textit{resolvent} moment generating function as 
	\begin{equation*}
		W_{1}
		^{0}(x) = \lim_{N \rightarrow \infty}\left \langle \tr\frac{1}{x-H}\right\rangle = \sum_{\ell=0}^{\infty}\frac{\lim_{N \rightarrow \infty} \langle \tr H^{\ell}\rangle }{x^{\ell+1}}.  
	\end{equation*}
	The resolvent is in fact the Stieltjes transform of the limiting eigenvalue density function:
	\begin{equation*}
		W^{0}_{1}(x) = \int_{\text{supp} \rho }\frac{\rho(y)dy}{x-y}.
	\end{equation*}
	Thus if one can compute the resolvent, then in theory one can invert the Stieltjes transform to find $\rho$.

	Spectral phase transitions are common phenomena involving  the limiting eigenvalue distribution and occurs when the number of connected components of the support changes as one changes the values of the coupling constants. We refer to Section \ref{spectral phase transition} for examples.
	
	For later use we will also define higher moments as 
	\begin{align*}
		\langle \tr H^{\ell_{1}} \tr H^{\ell_{2}}... \tr H^{\ell_{k}}\rangle = \int_{\mathcal{H}_{N}} \tr H^{\ell_{1}} \tr H^{\ell_{2}}... \tr H^{\ell_{k}} e^{-\frac{N}{2}\tr H^{2} - \sum_{j=3}^{d} \frac{N t_{j}}{j}\tr H^{j}}dH. 
	\end{align*}
	In practice however, it is often easier to compute the cumulants, denote with a subscript $c$, instead of the moments, which are related by the moment-cumulants, see relations chapter 1.2.5 of \cite{Eynard2018}.

	\section{Formal matrix models}\label{formal matrix models}
	We will now discuss a very different but deeply related type of matrix model. Informally, given an expression for a matrix integral that may or may not be convergent, one defines a \textit{formal matrix integral} by expanding the non-Gaussian terms in a power series and exchanging the order of integration and summation. For example, consider the expression
	\begin{equation*}
		\int_{\mathcal{H}_{N}}e^{-\frac{N}{2t}\tr H^{2} - \frac{t_4 N}{4t}\tr H^{4}}dH.
	\end{equation*}
	A formal matrix integral corresponding to this expression may be defined as 
	\begin{equation*}
		Z_{\text{quad}}:=\sum_{n=0}^{\infty}\frac{N^{n}t_{4}^{n}}{(4t)^{n}n!}\int_{\mathcal{H}_{N}}(\tr H^{4})^{n}e^{-\frac{N}{2}\tr H^{2}}dH.
	\end{equation*}
	In general, and as can be seen in the preceding example, formal matrix integrals are weighted formal summations of GUE moments. Note that these series are typically divergent and should be understood as formal power series in $t, t_4$ and other coupling constants.
	
\begin{figure}[H]
		\includegraphics[width=12cm]{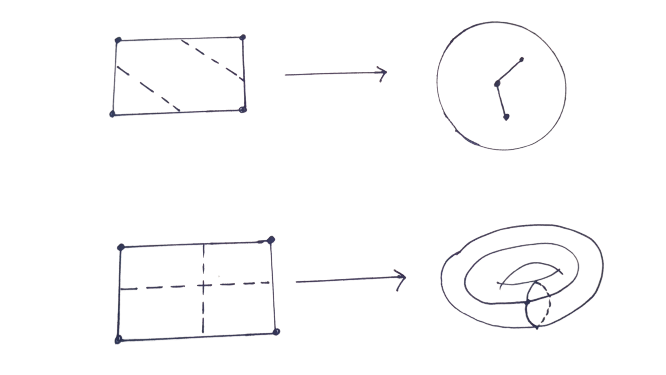}
		\centering
		\caption{The only two possible maps resulting from gluing an unmarked quandrangle.}
		\label{map example}
	\end{figure}

	 GUE moments have a combinatorial interpretation as sums over \textit{maps}. A map is a graph embedded in a Riemann surface that comes from gluing edges of various polygons in particular the complement of a map on a surface is a disjoint union of open discs. For a more detailed definition see \cite{Eynard2018}. For example consider Figure \ref{map example}. This leads to formal matrix integrals being the weighted generating functions of maps \cite{BIPZ,Eynard2018}.

	Consider the previous example. One can show that 
	\begin{equation*}
		Z_{\text{quad}} = \sum_{g \geq 0}\left(\frac{N}{t}\right)^{2-2g}\left[ \sum_{v=0}^{\infty}t^{v}\sum_{\Sigma\in \mathcal{M}^{g}(v)}\frac{t_{4}^{n_{4}(\Sigma)}}{|\text{Aut}(\Sigma)|}\right] 
	\end{equation*}
	where, for the quartic model, $\mathcal{M}^{g}(v)$ is the set of maps with genus $g$ and $v$ vertices formed from gluing quadrangles together and $n_4 (\Sigma)$ is the number of quadrangles glued to form the map $\Sigma$. One can show that for fixed $g$ and $v$, $\mathcal{M}^{g}(v)$ is a finite set, thus the coefficients of the series 
	\begin{equation}\label{free energy expansion}
		F_{g} :=  \sum_{v=0}^{\infty}t^{v}\sum_{\Sigma\in \mathcal{M}^{g}(v)}\frac{t_{4}^{n_{4}(\Sigma)}}{|\text{Aut}(\Sigma)|}
	\end{equation}
	are finite. This ensures that $Z_{quad}$ is a well-defined formal series  \cite{Eynard2018}.
	
	In general, the consequence of a term of the form
	\begin{equation*}
		\frac{t_{j}N}{j t} \tr H^{j}
	\end{equation*}
	in the potential is that one adds  $j$-gons to the set of  polygons that may be used to glue maps. Thus, a formal matrix integral of the form
	\begin{equation*}
		Z_N = \int_{\mathcal{H}_{N}}e^{-\frac{N}{2t}\tr H^{2} - \sum_{j=3}^{d}\frac{N t_j}{jt}\tr H^{j}}dH
	\end{equation*} 
	is in fact the formal power series
	\begin{equation}\label{gen function}
		Z_N = \sum_{g \geq 0}\left(\frac{N}{t}\right)^{2-2g} \left[\sum_{v=0}^{\infty}t^{v}\sum_{\Sigma\in \mathcal{M}^{g}(v)}\frac{t_{3}^{n_{3}(\Sigma)}t_{4}^{n_{4}(\Sigma)}... t_{d}^{n_{d}(\Sigma)}}{|\text{Aut}(\Sigma)|}\right],
	\end{equation}
	where  $\mathcal{M}^{g}(v)$ is the set of maps with genus $g$ and $v$ vertices formed from gluing:
	triangles, quadrangles, ..., and  $d$-gons. For $1\leq j \leq d$,  $n_j (\Sigma)$  is the number of $j$-gons used to glue the map $\Sigma$. This formal summation is once again well-defined because the set  $\mathcal{M}^{g}(v)$ is finite.
	
	Note that formal matrix integrals like (\ref{gen function}) are organized by the genus of the maps. Such an expression is called a  large $N$ expansion or \textit{genus expansion}. Remarkably, convergent matrix integrals often also have a genus expansion whose coefficients coincide to leading order with those of its formal counterpart. For discussions on the relationship between formal and convergent matrix models see \cite{Eynard combin}.
	
\begin{figure}[H]
		\includegraphics[width=12cm]{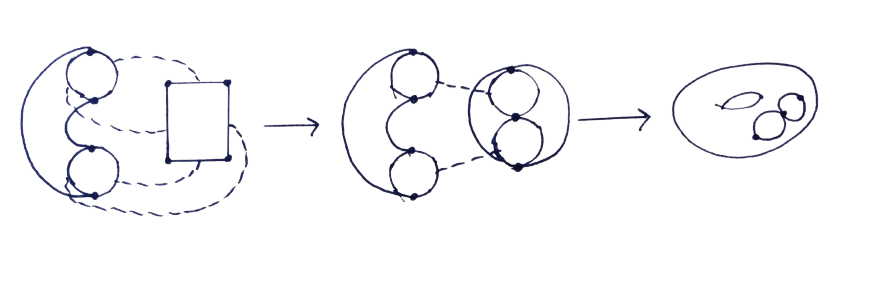}
		\centering
		\caption{A gluing resulting from a quandrangle and a 2-cell with the topology of a cylinder and boundaries of lengths two.}
		\label{map stuff}
	\end{figure}

	We will now extend these ideas to bi-tracial formal matrix models. A 2-cell of topology $(k,g)$ is a connected oriented genus $g$ surface with $k$ boundaries. They generalize the usual polygons seen in the perturbative expansion of single trace matrix models.   In particular, for bi-tracial matrix models, the consequence of a term of the form
	\begin{equation*}
		\frac{t_{j}N}{(i+j )t} \tr H^{i}\tr H^{j}
	\end{equation*}
	in the potential is that one adds 2-cells with the topology of a cylinder with boundaries of lengths $i$ and $j$ to the set of polygons to glue. Maps glued from 2-cells with topologies other than a disc are referred to as \textit{stuffed maps} and were studied in detail in \cite{blobbed1,blobbed2}. For example see Figure \ref{map stuff}.
	
	Some other important aspect of formal matrix integrals are their moments, higher moments, and cumulants. Consider the following formal matrix integral
	
	\begin{equation*}
		\int_{\mathcal{H}_{N}} e^{-\frac{N}{2t}\tr H^{2} - \sum_{j=3}^{d} \frac{N t_{j}}{jt}\tr H^{j}}dH.
	\end{equation*}
	Its moments are the formal series that result from considering the integral of $\tr H^{\ell}$ then power series expanding all non-Gaussian terms and swapping the order of integration and summation. One can show that these moments, just like the partition function, have a genus expansion:
	\begin{equation*}
		\langle \tr H^{\ell} \rangle  =\sum_{g \geq 0}\left(\frac{N}{t}\right)^{1-2g} \left[\sum_{v=0}^{\infty}t^{v}\sum_{\Sigma\in \mathcal{M}^{g}_{\ell}(v)}\frac{t_{3}^{n_{3}(\Sigma)}t_{4}^{n_{4}(\Sigma)}... t_{d}^{n_{d}(\Sigma)}}{|\text{Aut}(\Sigma)|}\right] := \sum_{g\geq 0} \left(\frac{N}{t}\right)^{1-2g}\mathcal{T}^{g}_{\ell}
	\end{equation*}
	where $\mathcal{M}^{g}_{\ell}(v)$ denotes the set of connected maps of genus $g$, with $v$ vertices, glued from triangles,  quandrangles, ..., $d$-gons, and one distinguished $\ell$-gon with a rooted edge, which is called a \textit{boundary}. By rooted edge we mean that one edge is distinct and has a direction that orients the polygon. Notice that maps in $\mathcal{M}^{g}_{\ell}(v)$ are connected, where by definition a map is called connected if it is draw on a connected Riemann surface.  Note that if $\ell = 0$, then each  $\mathcal{T}^{g}_{0}$ is precisely $F_{g}$ from equation (\ref{free energy expansion}).
	
	Next one wants to consider higher moments which have a similar genus expansion
	\begin{equation*}
		\langle \tr H^{\ell_{1}}\tr H^{\ell_{2}}\cdots \tr H^{\ell_{k}}\rangle  =\sum_{g \geq 0}\left(\frac{N}{t}\right)^{2-2g-k} \left[\sum_{v=0}^{\infty}t^{v}\sum_{\Sigma\in \mathcal{M}^{g}_{\ell_{1},...\ell_{k}}(v)}\frac{t_{3}^{n_{3}(\Sigma)}t_{4}^{n_{4}(\Sigma)}... t_{d}^{n_{d}(\Sigma)}}{|\text{Aut}(\Sigma)|}\right],
	\end{equation*}
	where $\mathcal{M}^{g}_{\ell_{1},...,\ell_{k}}(v)$ denotes the set of not necessarily connected maps of genus $g$, with $v$ vertices, glued from triangles,  quandrangles, ..., d-gons, one boundary of length $\ell_{1}$,..., one boundary of length $\ell_{k}$. Notice that this time the set is not necessarily connected. Counting connected maps is much easier than counting disconnected ones, thus in practice one computes the sum over connected maps, denoted by	$\langle \tr H^{\ell_{1}}\tr H^{\ell_{2}}\cdots \tr H^{\ell_{k}}\rangle_{c}$. Remarkably, the connected sums are precisely the cumulants from classical multivariate probability theory. In particular one can recover $	\langle \tr H^{\ell_{1}}\tr H^{\ell_{2}}\cdots \tr H^{\ell_{k}}\rangle$ using the moment-cumulant relations. For a discussion of this see chapter 1.2.5 of \cite{Eynard2018}. We will see in the next section that one can compute the coefficients of the genus expansions of moments and cumulants recursively.
	
In formal matrix integrals we have seen that the partition function, moments, and cumulants are well-defined formal series. It is often the case that these converge in some small multi-disc  while also having (usually algebraic) singular behaviour for certain values of the coupling constants. At these singularities, the formal series has an asymptotic expansion. The exponent of the leading order term in this asymptotic expansion is usually referred to as a \textit{critical exponent}. Let $A(x)$ be a formal series with a singularity at $x_{c}$, then the critical exponent of
\begin{equation*}
	A(x)\sim C(x-x_{c})^{a} + ...
\end{equation*}
is $a$. 

This idea is common in statistical mechanics to describe quantities near the singularities of phase transitions and are in some sense universal. In random matrix theory they are often used to find connections to areas of physics, such as conformal field theory \cite{Ds multi 2}. Some examples of this are given in Section \ref{Liouville quantum } for the genus expansion terms of the natural logarithm of the partition function.

	\section{The Schwinger-Dyson equations and Topological Recursion}\label{SDE's and TR}
	A common tool used to analyze both formal and convergent random matrix integrals are Schwinger-Dyson equations (SDE's): an infinite system of recursive equations between moments and cumulants. They were first introduced by Migdal in \cite{Migdal}.  These equations have a straightforward derivation. Consider a matrix integral, either formal or convergent, of the form 
	\begin{equation*}
		\int_{\mathcal{H}_{N}}e^{-S(H)}dH,
	\end{equation*}
	where $S(H)$ is some multi-tracial polynomial of powers of $H$. Using Stoke's formula, it follows that 
	\begin{equation*}
		\sum_{i,j=1}^{N}	\int_{\mathcal{H}_{N}} \frac{\partial}{\partial H_{ij}} \left((H^{\ell})_{ij} e^{-S(H)}\right)dH =0
	\end{equation*}
	for any $\ell \geq 0$. Applying the product rule to the integrand we find that 
	\begin{equation*}
		\sum_{k=0}^{\ell-1} \langle \tr H^{\ell-1-k} \tr H^{k} \rangle - \langle \tr H^{\ell}S'(H) \rangle  =0.  
	\end{equation*}
	
	Suppose for example that $S(H) = N/2 \tr H^{2} + N t_{4}/4 \tr H^{4}$. Then this equations becomes 
	\begin{equation}\label{quartic example}
		\sum_{k=0}^{\ell-1} \langle \tr H^{\ell-1-k} \tr H^{k} \rangle = N \langle  \tr H^{\ell+1} \rangle  + N \langle  \tr H^{\ell+4} \rangle.   
	\end{equation}
	
	Often finding the solutions for these equations at finite $N$ is quite difficult. By applying the genus expansion of moments, provided that it exists, one can derive equations  for each order of $N$ that relate the coefficients of the genus expansions. Continuing the quartic example, let
	\begin{equation*}
		\langle \tr H^{\ell}\rangle := \sum_{g \geq 0} N^{1-2g} \mathcal{T}^{g}_{\ell}.
	\end{equation*}
	
	Collecting like terms of the genus expansion of the moments in the Schwinger-Dyson equations, the leading order SDE's in $N$ that come from (\ref{quartic example}), often called the \textit{loop equations},   are
	\begin{equation}\label{quartic loop}
		\sum_{k=0}^{\ell-1}\mathcal{T}_{\ell-1-k}^{0} \mathcal{T}^{0}_{k} = \mathcal{T}^{0}_{\ell+1}  + \mathcal{T}^{0}_{\ell+4},
	\end{equation}
	for $\ell \geq 0$.
	This equation is much simpler to solve due to the disappearance of the higher moments. In particular, in the case of a Gaussian potential, all odd moments are zero, let $\ell = 2n +1$  for $k\geq 0$ then equation (\ref{quartic loop}) becomes
	\begin{equation*}
		\sum_{k=0}^{2n}\mathcal{T}_{2n-2k}^{0} \mathcal{T}^{0}_{2k} = \mathcal{T}^{0}_{2n+2} 
	\end{equation*} with $\mathcal{T}^{0}_{0} =1$. The solution is $\mathcal{T}^{0}_{2n} = C_{n}$ the $n$th Catalan numbers, which are the leading order terms in the genus expansion  of GUE moments.
	
	This is the main reason to introduce the genus expansion is that it allows access to simpler Schwinger-Dyson equations by restricting to leading order. However, it is  possible to recover lower order contributions through a process called Topological Recursion which we will outline now.
	
	Note that this process can be repeated for higher order moments by considering
	\begin{equation*}
		\sum_{i,j=1}^{N}	\int_{\mathcal{H}_{N}} \frac{\partial}{\partial H_{ij}}\left( (H^{\ell_{1}})_{ij} \prod_{q=2}^{m}\tr H^{\ell_{q}} e^{-S(H)}\right)dH =0.
	\end{equation*}
	Denote the genus expansion terms of general cumulants as 
	\begin{equation*}
		\langle \tr H^{\ell_{1}}\tr H^{\ell_{2}}\cdots \tr H^{\ell_{n}}\rangle_{c} := \sum_{g \geq 0} N^{\chi}\, \mathcal{T}^{g}_{\ell_{1},\ell_{2},...,\ell_{n}},
	\end{equation*}
	where $\chi = 2-2g-n$ is the Euler characteristics of the maps in that term of the expansion. Then, for the example of the quartic potential from before, one can find that 
	
	\begin{equation*}
		2\sum_{k=0}^{\ell_{1}-1}\mathcal{T}_{\ell_{1}-1-k}^{0} \mathcal{T}^{0}_{k,\ell_{2}} +\ell_{2} \mathcal{T}^{0}_{\ell_{1}-1+\ell_{2}} = \mathcal{T}^{0}_{\ell_{1}+1,\ell_{2}}  + \mathcal{T}^{0}_{\ell+4,\ell_{2}},
	\end{equation*}
	for $\ell_{1},\ell_{2}\geq 1$. From the terms in the product on the left-hand side this equation relies on the solution to equation (\ref{quartic loop}). Similarly,  each equation for a given Euler characteristic relies on the solutions of the higher Euler characteristic equations. For details on the SDE's see \cite{Random Matrices,Eynard2018} or in the case that there is a multi-trace potential \cite{blobbed1,AK,HKP 2}.

	In summary the SDE's are an unwieldy infinite set of recursive equations between terms of the genus expansion of moments or cumulants. However, in \cite{Eynard TR} a method was outlined  to streamline this process, called \textit{Topological Recursion}. This process has been generalized  \cite{Eynard TR summary,Airy structures,Quantum Airy structures} from its original use in matrix integrals. We start by defining the following generating functions of the genus expansion terms of moments and cumulants:
	\begin{equation*}
		W^{g}_{n}(x_{1},x_{2},...,x_{n}) = \sum_{\ell_{1},\ell_{2},...,\ell_{n}} \frac{\mathcal{T}^{g}_{\ell_{1},\ell_{2},...\ell_{n} }}{x_{1}^{\ell_{1}+1}x_{2}^{\ell_{2}+1}\cdots x_{n}^{\ell_{n}+1}}.
	\end{equation*}
	One can write the SDE's in terms of these new generating functions as a means of collecting terms.  In single or bi-tracial single Hermitian matrix models Topological Recursion allows one to compute any  $	W^{g}_{n}$ from just the information in the resolvent $	W^{0}_{1}$ and the \textit{cylinder amplitude} $W^{0}_{2}$. The cylinder amplitude is often in some sense universal. This means that the only fundamental information needed to compute any  $	W^{g}_{n}$  is contained in the resolvent. Once one has sufficient $	W^{g}_{n}$'s it is possible to compute the genus expansion terms of the partition function directly. For details on this process see chapter 3 of \cite{Eynard2018}. For single matrix integrals with higher trace multiplicities than two one needs a generalized process called Blobbed Topological Recursion \cite{blobbed1,blobbed2}.
\end{appendices}

\end{document}